\documentclass{aastex}
\usepackage{emulateapj5}

\newcommand{\ie}{i.e.,}
\newcommand{\eg}{e.g.,}

\newcommand{\etal}{et~al.\ }
\newcommand{\ltsima}{$\; \buildrel < \over \sim \;$}
\newcommand{\simlt}{\lower.5ex\hbox{\ltsima}}
\newcommand{\gtsima}{$\; \buildrel > \over \sim \;$}
\newcommand{\simgt}{\lower.5ex\hbox{\gtsima}}

\newcommand{\magsec}{mag~arcsec$^{-2}$}

\def\rquart{$r^{1/4}$~}
\def\parcmin{{\tt '}\mskip -6.0mu.\,}
\def\parcsec{{\tt ''}\mskip -6.0mu.\,}
\def\muv{$\mu_{\mbox{v}}$}

\begin{document}
 
\title{Deep CCD Surface Photometry of Galaxy Clusters II: Searching for
Intracluster Starlight in non-cD clusters}

\author{John J. Feldmeier\altaffilmark{1,2},  J. Christopher 
Mihos\altaffilmark{1,3,4}, Heather L. Morrison\altaffilmark{4}, 
Paul Harding, Nathan Kaib\altaffilmark{1,5}}
\email{johnf@bottom.astr.cwru.edu, hos@onager.astr.cwru.edu, 
heather@vegemite.astr.cwru.edu, harding@dropbear.astr.cwru.edu, 
kaib@astro.washington.edu}

\affil{Department of Astronomy, Case Western Reserve University,
10900 Euclid Ave, Cleveland, OH 44106}

\and
\author{John Dubinski}
\email{dubinski@astro.utoronto.ca}
\affil{Department of Astronomy and Astrophysics, 
University of Toronto, 60 St. George St., Toronto, ON M5S 3H8, Canada}

\altaffiltext{1}{Visiting Astronomer, Kitt Peak National Observatory, 
National Optical Astronomy Observatory, which is operated by 
the Association of Universities for Research in Astronomy, Inc. 
(AURA) under cooperative agreement with the National Science Foundation.}

\altaffiltext{2}{NSF Astronomy and Astrophysics Postdoctoral Fellow}

\altaffiltext{3}{Cottrell Scholar of Research Corporation and 
NSF CAREER fellow}

\altaffiltext{4}{Also Department of Physics, Case Western Reserve University}
\altaffiltext{5}{Current address: Astronomy Dept., Univ. of Washington, 
Box 351580, Seattle, WA 98195-1580}

\begin{abstract}

We report the search for intracluster light in four Abell Type II/III
(non-cD) galaxy clusters: Abell 801, 1234, 1553, \& 1914.  We find on
average that these clusters contain $\sim$ 10\% of their detected
stellar luminosity in a diffuse component.  We show that for two of
the clusters the intracluster light closely follows the galaxy
distribution, but in the other two cases, there are noticeable
differences between the spatial distribution of the galaxies and the
intracluster light.  We report the results of a search for
intracluster tidal debris in each cluster, and note that Abell~1914 in
particular has a number of strong tidal features likely due to its
status as a recent cluster merger.  One of the Abell~1914 features
appears to be spatially coincident with an extension seen in weak
lensing maps, implying the feature traces a large amount of mass.  We
compare these results to numerical simulations of
hierarchically-formed galaxy clusters, and find good general agreement
between the observed and simulated images, although we also find that
our observations sample only the brightest features of the
intracluster light.  Together, these results suggest that intracluster
light can be a valuable tool in determining the evolutionary state of
galaxy clusters.

\end{abstract}
 
 
\section{Introduction}

As the most massive gravitationally 
bound structures in the universe, galaxy clusters
stand to teach us much about the hierarchical assembly of matter in
the universe. Clusters exhibit a wide variety of structural
properties, from massive, X-ray luminous clusters dominated by early
type galaxies (i.e., Coma) to irregular, spiral-rich clusters like the
Ursa Major cluster, down to poor clusters and loose groups.
The fact that many clusters are still obviously in the process of
assembly can be seen via many tracers of substructure, such as X-ray
isophotes, gravitational lensing maps, and kinematic and spatial
substructure in the galaxy populations (see, e.g., ~reviews by
Girardi \& Biviano 2002 and Buote 2002)

A potentially powerful new tracer of the assembly history of clusters
is intracluster light, the diffuse starlight which permeates many
galaxy clusters. Once simply another curiosity of Zwicky (1951),
individual intracluster stars have been clearly detected in several
nearby galaxy clusters \citep{1996arna,ftv1998,durr2002,ipn3}, and
through deep imaging this diffuse light has been detected in many more
distant clusters \citep{uson1991b,vg1994,bern1995,gon2000}. From the
results to date, it is clear that intracluster light (ICL) is a common
component of galaxy clusters and contains between 10\% and 50\% of the
total stellar luminosity of the cluster, albeit with large
uncertainties due to the intrinsic low surface brightness of the
component (less than 1\% of the night sky background).

The study of intracluster light is now entering a new phase, focusing
on what can be learned about the evolution of galaxy clusters and
their member galaxies using the ICL. Since galaxy clusters form
hierarchically, and since the bulk of ICL production is believed to
occur due to tidal-stripping from galaxy interactions and from the
mean tidal field of the cluster
\citep{rich1983,miller1983,mer1984,gnedin2003}, the properties of the
ICL should be intimately tied to the dynamical evolution of clusters
(\eg ~Merritt 1984; Moore \etal 1996; 
Dubinski 1998; Mihos 2003; Napolitano \etal 2003;
Willman \etal 2004). As clusters dynamically evolve, the fractional
amount of ICL should increase if the cluster is isolated, 
and its spatial and dynamical structure should become well-mixed 
in the cluster potential well.  As additional groups of 
galaxies enter the cluster, the fractional amount of ICL
will briefly decrease, as the newer galaxies are initially 
unstripped, and then the fraction will increase as the additional 
galaxies suffer the effects of the cluster environment. 
The exact details of this overall evolution depend on 
the mass and accretion history of the cluster, tying the properties 
of the ICL directly to the cosmic
history of cluster assembly. 

The exact mechanisms of ICL production have implications for a number of
other galaxy cluster studies.  Ultra-compact dwarf galaxies 
(Drinkwater \etal 2003), the formation of S0 galaxies 
(Quilis, Moore, \& Bower 2000, and references therein) and tidal debris in
nearby clusters (Trentham \& Mobasher 1998; Gregg \& West 1998; 
Calcaneo-Roldan \etal 2000) may all be closely related to the
ICL phenomenon.  Since searches for individual intracluster stars 
in nearby galaxy clusters can
be influenced by metallicity effects (Durrell \etal 2002; Feldmeier \etal
2004), knowing the dominant progenitor population is critical to
avoid underestimating the true fraction of intracluster light.
However, observational constraints on the
ICL are still extremely poor, due largely to the scarcity of
quantitative measurements of the ICL in clusters, especially over a
range of cluster properties.
  
To address these questions, we have recently begun deep imaging of a
sample of galaxy clusters to quantify the structure of ICL as a
function of galaxy cluster properties (Feldmeier \etal 2002; hereafter
Paper~I).  We have observed two cD-dominated (Bautz-Morgan Type
I) galaxy clusters thus far (Abell~1413 and MKW~7); in each we
quantify the extended cD envelope and find an excess of luminosity
over a pure \rquart law. However, this extended light is very smooth,
and we find relatively little substructure in the ICL of either
Abell~1413 or MKW~7.  Again, however, both of these clusters are
cD-dominated, and there are a number of reasons to expand our sample
beyond cD-dominated galaxy clusters.

First, less than twenty percent of the total number of Abell clusters
have a Bautz-Morgan type of I \citep{leir1977}.  Observations of ICL
in Type~I clusters may therefore not be representative of ICL
structure in galaxy clusters as a whole.  Second, the relation between
intracluster light and cD envelopes is still unclear.  Does the
presence of a cD galaxy in a cluster always imply a large amount of
intracluster starlight, or are ICL properties less correlated to the
precise details of the cluster core?  If intracluster stars are
predominantly removed early in the cluster's dynamical history (Merritt
1984), we would expect the properties of ICL to be intimately tied to
the process of cD formation.  If instead the bulk of the
intracluster stars are removed from their parent galaxies by late
tidal-stripping and related ``galaxy harassment'' scenarios operating
on low luminosity galaxies (Richstone \& Malumuth 1983; Moore \etal
1996), then the ICL properties may be less sensitive to the presence
or absence of a cD galaxy in the cluster core.  Finally, given the
expected evolution of galaxy clusters under hierarchical structure
formation models (\eg\ Merritt 1985; Dubinski 1998), it is expected that
most rich galaxy clusters will eventually form cD galaxies, though the
timescales involved could be quite long (up to thousands of Gyr; 
Adams \& Laughlin 1999).  Therefore Bautz-Morgan type
II or III clusters may be less dynamically evolved proxies of the more
evolved type I clusters, and ICL studies of these clusters may give us
insight into to the evolution of galaxy clusters at higher redshift
and younger dynamical ages.

\section{Cluster Properties}

For this particular study, we focused on Abell clusters of
Bautz-Morgan \citep{bm1970} type of II, or greater, and similar
richness (richness class 2).  For our general survey goals and
detailed selection criteria, see Paper~I.  Briefly, we focus on
galaxy clusters that are smaller than the field of view of the
telescope/detector combination (in this case, the KPNO 2m + TK2A chip;
see \S3), taking care to avoid nearby bright stars or galaxies that
could be a significant source of scattered light.  Table~1 gives the
coordinates and basic information for each galaxy cluster observed in
this study.  Here, we briefly summarize the relevant properties for
each cluster.  In particular, we emphasize the known X-ray properties of
our clusters, since X-ray detections are one of the most reliable
indicators that a galaxy cluster is bona-fide, and X-ray observations
give good indications of the mass and overall structure of a cluster.

Abell~801 has a Rood-Sastry type of B (b) \citep{sr1987}, a binary cluster 
where the central binary pair is connected with an optical bridge.  It 
has been detected in X-rays multiple times \citep{soltan1983,noras2000}, 
but has no measured X-ray temperature.  Baum (1973) 
using photoelectric data, 
observed Abell~801 for intracluster light, and found that the intracluster
light was $\approx$ 16\% of the total cluster light, defined as the 
light outside of the \muv = 26 isophote of the cluster galaxies.  
By extrapolating the luminosity function of Abell (1962), 
Baum claimed that approximately half of this excess could be 
attributed to dwarf galaxies below the Palomar Sky Survey
plate limit, bringing the total amount of intracluster light to 
$\approx$ 8\%.  Gudehus (1989) claimed that the remaining
excess light could be accounted for by the presence of dwarf 
galaxies below the surface brightness threshold, and the overlapping 
portions of bright galaxies, but gave no quantitative 
evidence to support this claim.

Abell~1234 has a Rood-Sastry type of L (line), where four or more of 
the ten brightest galaxies are arranged
approximately collinearly with numerous fainter members located
around them \citep{sr1987}.  It has been detected in the X-ray 
\citep{burns1994,dfj1999} and significant X-ray substructure
has been observed \citep{burns1994}, but no X-ray temperature
has been determined.  Surface photometry of a radio galaxy
in the core of this cluster has been obtained \citep{ledlow1995}, 
allowing us to compare our surface photometry on these spatial
scales with others (\S 5.4).
Finally, \citet{sr1987} note that this cluster is 
connected to Abell~1246.

Abell~1553 also has a Rood-Sastry type of L. From galaxy counts,
\citet{bucknell1979} note that the entire cluster is considerably
elongated, and \citet{yama1986} claim that the cluster is quite large,
with a cluster radius as large as R=11$\parcmin$0 (1.75 Mpc),
though the measured core radius is much smaller R=2$\parcmin0$ (300
kpc).  \cite{sr1987}   note that this cluster appears 
closer than indicated for its distance class, implying 
possible foreground galaxy contamination.  Abell~1553 has 
been detected in X-rays numerous times
\cite[\eg][]{bohr1994,noras2000,shibata2001}.  \citet{white2000}
derived a temperature of 4.18 $\pm$ 0.15 keV for Abell~1553 from {\sl
ASCA} data, but this temperature assumes a single thermal component
model, and the temperature profile in the {\sl ASCA} data is dropping
significantly with radius, possibly due to poorer quality data 
or problems with background subtraction.  The
latter is more likely to be the case: Abell~1553 is directly behind
the nearby Virgo cluster of galaxies.  The center of Abell~1553 is
$\approx$ 1.8 degrees to the south of M~87, and is clearly surrounded
by X-ray emission from the foreground cluster
\citep{bohr1994,shibata2001}.  Virgo has an average X-ray temperature
of $\approx$ 2.5 keV \citep{shibata2001}, therefore it is likely that
Abell~1553's true temperature is larger than previously derived.  The
possible effect of Virgo's foreground light to our ICL observations 
will be discussed in \S 5.5

Finally, Abell~1914 also has a Rood-Sastry type of L.  Its X-ray
structure was claimed to be regular and smooth (Buote \& Tsai 1996),
but Jones \etal (2001) subsequently showed that X-ray images showed
significant substructure in the core, that a conventional $\beta$
model did not fit the data well, and that the cluster was in the
process of merging.  Abell~1914 has a high X-ray temperature
\citep[10.53$^{+0.51}_{-0.50}$ keV;][]{white2000}, although
Ikebe \etal (2002) find a somewhat lower
temperature of T = 8.41$^{+0.60}_{-0.58}$ keV.  Abell~1914 also has a
measured Sunyaev-Zeldovich decrement \citep{grego2001,jones2001},
which with the high X-ray temperature, suggests that this cluster may
be more massive than the other three in this study.  Abell~1914 is one
of approximately thirty galaxy clusters that are currently known to
contain radio halos and relics \citep{gio1999,kemp2001}, which are
thought to be the result of cluster mergers (\eg  ~Feretti 1999; Buote
2001).  Abell~1914 was analyzed using weak lensing by
\citet{dlens2002}, who note that the light distribution and the weak
lensing signal are highly elongated toward the south and west,
supporting the merging state of this cluster.  They also note the
presence of several red and blue gravitational arc candidates around
the brightest cluster galaxy.

\section{Observations}

The data for all four clusters were obtained over a seven night dark run
in 2002 March 11--17, using the 2.1m telescope at Kitt
Peak National Observatory\footnote{Kitt Peak National Observatory is a
division of NOAO, which is operated by AURA, under cooperative
agreement with the National Science Foundation.}.  The images were
taken using a 2048 x 2048 Tektronix CCD (T2KA).  With this setup, the
field of view was 10.4 arcmin$^{2}$, with each pixel imaging
0.305\arcsec of sky.  The gain was set at the default value of 3.6
e$^{-}$~ADU$^{-1}$ and the readout noise was 4~e$^{-}$ (1.1 ADU).  All
exposures were made through a Washington {\it M} filter (see Paper~I
for a discussion on filter choice).  We transformed these observations
to Johnson {\it V} (\S 4.1), and unless otherwise stated, all surface
brightnesses in this paper are given in {\it V} magnitudes.  The first
two nights of the run had photometric conditions, and excellent seeing
($0\parcsec8$ - $1\parcsec0$).  On these nights, we observed
Abell~1234 and Abell~1914.  Unfortunately, before the third night, a front
passed overhead, degrading the seeing 
to an average of $2\parcsec0$,
though conditions were still photometric for the next two nights.  We
observed Abell~801 and Abell~1553 under these degraded conditions.

An accurate flat-field is critical to the success of deep surface
photometry such as ours.  As
mentioned in \S 1, we are interested in recovering a signal that is
less than 1\% of the sky background.  Our flat field must be at least
a few times more accurate than this 1\% value for our data to be
meaningful.  For this reason, dome flat fields cannot be used due to
possible scattered light, differing pupil illuminations, and intrinsic
color differences.  For similar reasons, twilight flats are also
inadequate for our purposes.  Therefore, dark sky flats are a
necessity, and we obtained these flats in the manner described by
Morrison \etal(1997).  Half of the time was used observing the galaxy
clusters, and the other half was used to obtain dark sky flats to
guard against systematic errors due to telescope/detector flexure.  
The dark sky flat images were taken at pre-determined areas away from
bright stars at approximately the same hour angle and declination as
the cluster images (see Table~2).  For each of the cluster and sky
images, the exposure time was 900~s.  We dithered the cluster images a
few arcseconds between exposures to average out pixel-to-pixel
flatfield variations, and we dithered the sky flat images by at least
five arcminutes to avoid the overlapping wings of bright stars.

\section{Data Reduction}

The data reduction procedures for this survey 
were given in detail in Paper~I: we briefly summarize them here.  
Overscan removal, bias subtraction and trimming of the images
were done in the usual manner using IRAF\footnote{
IRAF is distributed by the National Optical Astronomy Observatories,
which are operated by the Association of Universities for Research
in Astronomy, Inc., under cooperative agreement with the National
Science Foundation.}.  Following the procedures of Paper~I, we then
corrected our data for the known non-linearity of the T2KA detector 
(Mochejska \etal 2001), using the following model: 

\begin{equation}
I_{e} = I_{i} \cdot (c_{1} + c_{2} \cdot \frac{I_{i}}{32767} + 
c_{3} \cdot (\frac{I_{i}}{32767})^{2})
\end{equation}  
where $I_{i}$ is the measured intensity, and $I_{e}$ is the corrected
intensity in ADU.  The constants were determined by multiple linearity
tests using dome flats taken during our run with a constant lamp
voltage, but varying exposure time.  We interleaved a series of
one second exposures to track any change in the mean flux of the
dome flat lamp.  We found the following constants:  

\begin{equation}
c_{1} = 0.985 \pm 0.005, c_{2} = -0.297 \pm 0.002, 
c_{3} = 0.158 \pm 0.002
\end{equation}  

Note that the $c_{1}$ value is in good agreement with the results
of Mochejska \etal (2001), but the other two values are significantly
larger in amplitude.  This may be due to further changes with the T2KA
detector over time.  The actual correction to our surface brightness 
measurements will be extremely minor, because the intracluster light
has a low flux (see Figure~3 of Paper~I).  Nevertheless, the uncertainty
in our linearity correction is added to our final error model (see \S 6.4).

We next constructed a ``master'' sky flat from the 49 dark sky
images taken.  First, each individual sky flat was visually inspected
to ensure that no bright stars or scattered light patterns were
present in the image.  In four of the images, we found a strong linear
scattered light feature.  We created a mask for each
of the affected images to ignore these features, and a padding of 
$\approx$ $30\arcsec$ surrounding them.  No scattered light patterns were 
seen in any of the cluster images.  Once the scattered light patterns 
were masked, the flat was then constructed using the iterative procedure 
described in detail in Morrison \etal (1997) and in Paper~I.  We
correct for sky variations that are present in each image due to
airglow \citep{roach1973,wild1997,zheng1999} by fitting and 
dividing a normalized plane from each image.  
The entire process was iterated fifteen times, until the modes
of each individual sky image were well determined.

The galaxy cluster images were then flat-fielded by this final flat,
and were registered using stars common to all frames and the IRAF 
tasks GEOMAP and GEOTRANS, using a 2nd order polynomial fit.  
A preliminary sky value was found for each cluster image by finding
the mode of two regions on each chip well away from the 
center of the cluster, and averaging the results.  
This sky value was then subtracted from each image.  
The median sky value for each cluster is given in Table~2, column 3, 
and correlates
in the expected manner with lunar phase and time of night 
\citep[][and references therein]{kris1991,kris1997}.
After applying the photometric zero point in (\S 4.1) below, these 
sky values range between \muv ~= 21.61 and \muv ~= 21.12,
in agreement with the expected influence of the moon.  
Since the source of sky brightness is mostly  
within the earth's atmosphere, we removed our 0.17 mag~airmass$^{-1}$
extinction correction for our images around new moon, and found that 
the average brightness of the night sky at zenith near new 
moon was approximately \muv ~= 21.78.  This is in good
agreement with the measurements of \citet{kris1997}, who found a
surface brightness of $21.77 \pm 0.12$ at approximately the 
same time in the solar cycle (two years past the solar maximum). 

With the overscan, bias-subtraction, flat-fielding and sky 
subtraction complete for each image, we then combined the 
images together, using a 2$\sigma$ clipped median as before, and 
scaling for airmass.  We trimmed off portions of the combined image 
that were not in common between all of individual frames, which makes 
the final image for each cluster somewhat different in size (see Table~2
for final image sizes). The final images for each cluster are 
displayed in Figure~\ref{fig:clusters}.  

\subsection{Photometric Zero Point:}

The Landolt star fields SA~98, SA~107, SA~110, PG1633+099, Rubin~149,
\& Rubin~152 (Landolt 1992) were observed, giving us a total of 51
well-observed standard stars over a range of color and airmass. For
the purposes of our analysis, we converted our Washington {\it M}
exposures to {\it V} band magnitudes during the photometric
calibration process.  A photometric zero point of \muv ~= 21.09 $\pm
0.06$ (corresponding to 1~ADU~s$^{-1}$~pixel$^{-1}$, and assuming a
(B--V) color of 1.0) at unit airmass was determined.  For a 900 s
exposure this yields \muv ~= 28.48 corresponding to 1~ADU per pixel
at unit airmass, virtually identical to the zero point determined in
Paper~I.  As our exposures were only taken in one filter, and we do
not know the exact color of the intracluster light, we cannot add a
color correction term to our target photometry, but from the standard
star observations, we estimate its magnitude as less than 0.1 mag,
over the entire likely color range of our target objects (0.8 $\leq$
(B--V) $\leq$ 1.3).  The color term is reasonably well fit as a
linear function of (B--V), with a slope of 0.2 magnitudes per
magnitude of (B--V) color.

\section{Masking and Determination of Errors}

We adopt approximate size distances to our galaxies in 
Table~2, column 7, assuming the redshifts given in \citet{sr1999}, a Hubble 
constant, H$_{0} = $ 75~km~s$^{-1}$~Mpc$^{-1}$, and a cosmology 
of $\Omega_{\mbox{m}} = 0.3$, $\Omega_{\Lambda} = 0.7$.  At these 
small redshifts, these distances depend little on $\Omega$.  Given 
these assumed distances, we give the angular scale of each of
our cluster images in Table~2, column 8.  The corresponding luminosity 
distance moduli, ignoring any K-corrections, are given in Table~2, 
column 9.

\subsection{Masking}

In order to reach the faintest possible surface brightness levels of 
the intracluster light, we must mask out all other sources --- 
both stars and galaxies --- in the frame.  This allows us to be confident
that we are measuring the true sky values of each image, and not 
compromising our data analysis with model-dependent assumptions from
subtracting stars and galaxies.  We mask the objects by 
creating a binary mask image for each cluster where good pixels
are assigned a value of 1 and masked pixels a value of 0.

We first began by masking out the stars in each cluster image.  Since we are 
concerned with a very low surface brightness signal, we determined 
the point spread function (PSF) out to very large radii.  
Using the DAOPHOT photometry package (Stetson 1987), we detected all of
the stars in each cluster image down to a signal-to-noise of three, 
and used a subset  of bright stars to determine the PSF in each cluster
out to a radius of 20 pixels.  We then used this preliminary PSF to 
mask out all of the stars and small galaxies around the brightest 
saturated star in each cluster image.  Saturated stars have much 
higher signal-to-noise in the far wings on the PSF, which are our 
primary concern here.  Other sources near the saturated star 
on each image, such as resolved galaxies and stellar diffraction 
spikes, were removed manually.  
Then the unmasked pixels from the each saturated star were 
averaged in radial annuli, and joined to the smaller-radius 
PSF (which measures the inner core of the star more accurately).  
We then compared the four independent PSFs, one from each cluster image, 
which are displayed in Figure~\ref{fig:psf}.

Since the structure of PSF at large radii is due to effects such as
atmospheric aerosols and microripples and dust on the telescope optics 
(Racine 1996; Beckers 1995), it should be relatively insensitive 
to the exact values of the telescope seeing.  This is confirmed
in our data:  the four independent profiles taken under
varying seeing conditions agree quite well at intermediate radii.
At still larger radii, the signal-to-noise of the profiles approach
unity, and the profiles deviate from each other at a specific radius, 
which depends on the apparent magnitude of the saturated star.  
We therefore adopt the profile that extends securely 
to the largest radii, Abell~1234, and  we join that large-scale surface
brightness profile to the smaller scale profiles of the other clusters
at a radius of 12.2 arcseconds.  Using these large-radius 
PSFs, and the list of stars found by DAOPHOT, we masked all 
stellar sources in each cluster image out to a radius where the
magnitude-scaled PSF was 1 ADU above the sky value.  

Next, we masked out all of the galaxies in each cluster.  
Since unresolved galaxies would have already been treated as point sources, 
and thus been masked by the DAOPHOT procedure above, 
but many resolved sources remain in each cluster.  As in Paper~I, 
we masked these sources using the segmentation image from the 
SExtractor software package \citep[V2.2.2;][]{sex1996}, using 
identical parameters and the correction algorithms we used in
Paper~I to improve the masking.  Finally, the images multiplied 
by the mask were visually inspected, and any regions that needed 
further masking were masked using IMEDIT.  These manual 
corrections were usually due to ``islands'' of unmasked pixels 
(which have sizes of a few pixels and are discussed in detail 
in Paper~I), and were typically less than 0.3\% of 
the total image area.   

\subsection{Final Sky Subtraction, and Large-scale Flat-Fielding 
Errors}

Accurate sky subtraction is crucial to determine the true amount of
intracluster starlight in each cluster, and is one of the dominant
sources of error in our analysis.  We found a more accurate 
sky level for each cluster by using the masked image.  We first binned 
up each cluster image into bins of 49 $\times$ 49 pixels.  
For each bin, we calculated a robust average \citep{sb1994}, 
ignoring all masked pixels.  We then fit and subtract a plane from 
each masked, binned cluster image, using the IMSURFIT task in 
IRAF, taking care to use regions on each image that are well away 
from the cluster core.  We determined the locations of these
regions by first excluding any pixel within two arcminutes
of radius of the cluster core ($\approx$ 340kpc).  Then, we 
examined the binned image visually to find regions that were 
relatively free of any suspected ICL and bright stars and galaxies, 
to avoid any low-surface brightness emission that 
remained unmasked.  The average radii for these sky subtraction 
regions are typically 1 Mpc.  The mean corrections from this plane 
subtraction are small: less than 1.0 ADU for 
the clusters on average.  However, we emphasize that this process 
will remove any smooth ICL that covers the entire image.  
We then re-create the histogram of sky values in 49 $\times$ 49 pixel 
bins well away from the center of each 
cluster.  We also require that the bins contain at least 100 
unmasked pixels to be included in this revised histogram.  
These histograms are displayed in Figure~\ref{fig:hist}.  The
width of the histograms provides a measure of our uncertainties due to
large-scale flat-fielding errors and the faint outer wings of stars 
and galaxies that remain unmasked, even after the involved procedure
above.  

For Abell~801 and Abell~1234, the sky histograms are well-defined, with
a single mode, and relatively narrow distribution.  We adopted
the standard deviation in the sky histogram as the large-scale 
flat-fielding error in these two cases.  We note that the
values derived ($\approx$ 0.6 ADU) are noticeably smaller than the values 
derived for our first two galaxy clusters in Paper~I ($\approx$ 1.0 ADU), 
which we attribute to the larger number of sky flats available for
this data set (49 vs 20).  However, for Abell~1553 and Abell~1914, 
the sky histograms are clearly of lesser quality.  There is a clear 
bi-modality present in both histograms, and the width of the histograms
is much larger than the other two clusters.  It is important 
to note that the histogram range displayed in both of these clusters 
is $\pm$ 3 ADU from the sky level, which corresponds to a surface 
brightness of \muv ~= 27.2, or 6.0 magnitudes below the sky level.
Nevertheless, the effect is a systematic error of the data, and cannot
be ignored.  The sky histograms show a 
clear pattern along the y-axis of the image, corresponding to 
an E-W direction on the sky.  Figure~\ref{fig:flaterror} shows
this systematic effect for both Abell~1553 and Abell~1914 sky bins.  

Since all four clusters were flat-fielded with the same sky flat (\S 4),
this effect must be due to some systematic difference between 
the two groups of clusters (801 and 1234 versus 1553 and 1914).  
There are no clear differences between the two groups in seeing, 
lunar phase, or time of night observed.  However, we do note that
the hour angle that the clusters were observed  
is slightly different between the two groups.  In Table~2, column 5, 
we give the mean hour angle of the observations of each cluster.  
Abell~801 and 1234 tended to be observed east of zenith, while 
Abell~1553 and 1914 were observed slightly west of zenith.  The 
sky frames also have an average hour angle that is slightly east of 
the zenith.  Therefore, we attribute the large-scale flat-fielding
error to flexure in the telescope and detector 
as it tracks across the sky, and small mismatches between the hour
angle of our observations and the sky frames.  These small mismatches
would cause a slightly different vignetting pattern, and hence a
different flat field image.    

If the large-scale flat-fielding error is indeed due to the
telescope's hour angle, we could in principle partially correct this
effect to by creating two sky flats, one for the eastern portion of
our data and another for the western.  However, the number of sky flat
images on either side of zenith are unequal, and we would therefore
create a noisier flat for one half of our data.  In effect, this
approach would have us simply trading one source of flat-fielding
error for another in our analysis.  The large-scale flat-fielding
error could also be corrected by fitting a function to the sky bins
shown in Figure~\ref{fig:flaterror}, and dividing the data by this
function. However, this approach introduces a new source of
uncertainty from the functional fit to this higher-order structure in
the flat field.  This is particularly worrisome since our clusters are
always centered on the images, and may bias the fit in such a way to
artificially produce (or suppress) features which mimic the
large-scale ICL we are trying to detect.  We therefore chose the most
conservative approach -- to not attempt any higher-order corrections
to the flat field -- and adopt a large value for the large-scale
flat-fielding error for Abell~1553 and 1914 based on a visual
inspection of the uncorrected sky histograms. Although this limits the
surface brightness depths we reach in these clusters, we believe this
solution to be one which is least likely to bias our quantitative
measurement of the ICL in these clusters. The adopted large-scale
flat-fielding errors for all of the clusters is given in Table~3,
columns 1--4, where we note that even in the worst case the
flat-fielding error is smaller than 0.35\%.

\subsection{Limits to Our Precision}

The flux error model is described in detail in the Appendix, and is
used in all of our analysis.  At low surface brightnesses, the
dominant source of error are large-scale flat-fielding and
sky subtraction errors, which are systematic, and do not depend 
on the number of pixels averaged.  
Using our error model applied to the sky bins, we 
determined the error limit for each cluster
image in ADU, which we present in Table~3, column 5.  This limiting
uncertainty for each cluster can be converted into a surface
brightness limit.  Many authors prefer to quote the 1$\sigma$ limit as
their observational limit, and we give that limit, $\mu_{1\sigma}$,
for our cluster observations in Table~3, column 6.  These values range
between \muv=28 and 29.  However, in order
to be conservative, we will sometimes use a signal-to-noise ratio of five
in our analysis.  That limit, $\mu_{5\sigma}$, is given in Table~3,
column 7, and range between \muv=26 and 27, respectively.  
As an independent check of our error model, we compared the
expected standard deviation of the sky background from the error model
against the average standard deviation of the sky in each cluster
image found by SExtractor.  We found that our error model actually 
overestimated the measured standard deviation by 10--20\%: we 
attribute to this to the large-scale flat-fielding fluctuations 
seen in the data, but treated as sky variations in the 
SExtractor determination.  We are confident that our flux error 
model is a good description of the data.

\subsection{Comparison to previous results}

As a consistency check, we compare our surface photometry to other
published results from our clusters.  Unfortunately for the
clusters in this study, the only galaxy that has published
surface photometry results is 1119+216 (A1234-A; 
$\alpha$ = 11h 22m 29.95s, $\delta$ = +21\degr 24\arcmin 21\arcsec, J2000.0),
a radio galaxy studied by Ledlow \& Owen (1995; see Owen, White \& Ge 1993
for a finding chart).  The surface photometry of Ledlow \& Owen (1995) was
taken through a Cousins (R$_{\mbox{c}}$) scale, so we adopt the 
identical color transformation given by these authors (V--R$_{\mbox{c}}$
= 0.59) for comparison purposes.

We determine the surface brightness profile of 1119+216 
using the ELLIPSE task in IRAF/STSDAS \citep{busko1996}, based 
on the algorithms of 
\citet{jedr1987}.  We cannot compare the surface brightness profiles in
detail, since they are not published in Ledlow \& Owen (1995).  Instead
we compare our ELLIPSE results against theirs at r$_{24.5}$, 
where the R$_{\mbox{c}}$ surface brightness reaches 
24.5 \magsec.  The comparison is shown in Table 4.  The 
distances are computed assuming a H$_{0} = 75$~km~s$^{-1}$~Mpc$^{-1}$, 
q$_{0} = 0$ cosmology, to stay consistent with Ledlow \& Owen (1995).

There is good agreement between the parameters derived from both
datasets.  In the case of the position angle derived, we find
from our data that at the radius corresponding to r$_{24.5}$
is located within a 10 degree elliptical 
isophote twist, making it difficult to measure the position angle 
precisely.  We conclude that within the errors, our surface 
photometry for Abell~1234 is in agreement with previously 
published work, though our observations reach to a much fainter
surface brightness.

\subsection{Abell~1553 and Virgo's ICL}

As we have previously noted, Abell~1553 lies behind the the Virgo
cluster, which is known to have intracluster light of its own
\citep{1996arna,ipn1,ftv1998,durr2002}.  Does the presence of this
foreground intracluster light affect our measurements of the ICL of
Abell~1553?  Conversely, can we learn anything about the foreground
Virgo ICL from our observations of Abell~1553?

Since we fit and subtract a plane from each individual cluster
image before averaging the images together, any smooth ICL from the
Virgo cluster in our Abell~1553 frames will have been removed along
with the sky emission.  Due to the large fluctuations (10--30\%) in
the modal sky value due to atmospheric effects such as airglow, it is
almost impossible that we can directly detect any constant excess 
emission that we can attribute to Virgo's ICL.

However, we can set an interesting limit on the amount of intracluster
light in the region in front of Abell~1553.  Virgo's ICL is close
enough that any emission from it should also have surface brightness
fluctuations.  Tonry \& Schneider (1988) originally showed that the
variance (in ADU) of the mean surface brightness of a stellar population 
due to luminosity fluctuations
(their eq. 10) is:
\begin{equation}
\sigma^{2}_{L} = \overline{g}(x,y) t (10 \mbox{pc}/d)^{2} 
10^{-0.4(\overline{M}-m_{1})}
\end{equation}   
where $\overline{g}(x,y)$ is the mean sky-subtracted signal from the
source, $t$ is the exposure time of a single observation, $d$ is the
distance to the stellar population, $\overline{M}$ is the mean,
luminosity-weighted absolute magnitude of a stellar population and
$m_{1}$ is the magnitude of 1~ADU~s$^{-1}$~pixel$^{-1}$.  Normally,
this equation is used to find $d$, given a measurement of
$\sigma^{2}_{L}$, and $\overline{g}(x,y)$.  Here, our goal will be to
place a firm upper limit on $\overline{g}(x,y)$, assuming a distance
to the Virgo cluster, and a measurement of $\sigma_{L}$.

To obtain a firm upper limit to $\sigma_{L}$ on a large spatial scale, 
we proceed in the
following manner.  Since our error model already fully describes the
data (\S 5.3), to take a reasonable limit, we assume that the intrinsic
surface brightness
fluctuations have been inadvertently folded into the large-scale
flat-fielding error.  To separate the two effects, we take the
measurements of Abell~1553's large-scale flat-fielding error presented
in Figure~\ref{fig:flaterror}, and fit and subtract a quadratic
function through a least squares fit. 
After subtracting the quadratic, we find the standard deviation of the points
using a $3
\sigma$ clip of 0.670~ADU.  This is only slightly higher
than the minimum large-scale flat fielding error we derived for
Abell~801 (0.6 ADU). When we remove the minimum flat-fielding error
expected in quadrature, we are left with a value for $\sigma_{L} =
0.297$~ADU. We next need to apply a correction for the effects of
seeing, which artificially suppresses the intrinsic surface brightness
fluctuations. We obtain this correction factor by taking a simulated 
image with
mean zero and known $\sigma_{L}$, and convolving it with the measured
PSF (\S 5.1). The PSF-convolved image had a measured $\sigma_{L}$
which was a factor of 1.67 lower than the true value; we apply this
correction factor to our measurement of Abell 1553 to get a final
value of $\sigma_{L} = 0.497$~ADU. Since this represents the maximum
value possible for the intrinsic SBF of the Virgo ICL, it is clear
that there will be little or no effect on our studies of the ICL in
Abell~1553 itself on large spatial scales.

Using the relationship between SBF and surface brightness, we can now
place limits on the surface brightness of the Virgo ICL from our
data. After assuming a distance to Virgo of 15 Mpc, the photometric
zero point $m_{1} = 21.092$, and a $\overline{M_{V}}$ value of 1,
characteristic of old stellar populations \citep{tonry1990}, we find
that our upper limit value for $\overline{g}(x,y)$ is 5.7~ADU,
corresponding to a surface brightness of \muv ~= 26.6.  This surface
brightness limit is in rough agreement with values derived from direct
observations of Virgo's intracluster stars (\muv ~$\approx$ 27--29 ; 
Durrell \etal 2002;
Feldmeier \etal 2004), but is not faint enough to constrain the amount
of Virgo ICL significantly. Nonetheless, in the future, surface
brightness fluctuations may provide an important independent
constraint on the ICL in nearby clusters.

\section{Measuring the intracluster light}

The total cluster starlight in the galaxy clusters studied in Paper~I 
was clearly dominated by the central cD galaxy and the intracluster
light.  This made analysis reasonably straightforward: we masked the
other galaxies of the cluster, and treated the cD galaxy and the ICL
as a single, elliptically symmetric, luminous component.  However, in
the clusters studied here, the situation is more complicated.  There
is no central cD galaxy dominating the cluster, and the
intracluster light may not be elliptically symmetric (or even
present).  With the bulk of the cluster's optical light split evenly
between several galaxies, it becomes even more difficult to separate
the galaxy light from the intracluster light in a clear,
model-independent fashion.  Therefore, we have chosen to analyze these
clusters in several different ways in an effort to place reasonable
limits on the amount and spatial distribution of ICL in each cluster.
We unmask the light coming from galaxies, leaving only the point sources
masked in each galaxy cluster.

\subsection{Isophotal Measurements}

We first separated the optical light from galaxies and the ICL using
the simplest method possible: through an isophotal cutoff. The choice
of a unique isophotal cutoff is somewhat problematic,
however. Galaxies are often characterized by relatively high surface
brightness, and an common definition of the size of a galaxy is
$R_{25}$, the radius at which the surface brightness drops to
$\mu_B=25$ \citep{rc3}.  In reality, however, this tends to
underestimate the sizes of galaxies, which often extend to or exist at
much lower surface brightnesses \citep[\ie][]{sb1994,lsb1997,sb1999}.
Another benchmark of the surface brightness at which we might start to
define the intracluster light is to look at the surface brightness of
tidal debris in nearby interacting systems. For example, the merging
system NGC 7252 has two tidal tails with surface brightnesses $\mu_V
\sim 26-27$ (Hibbard \etal 1994, assuming B-V $\sim$ 1).  As there is
no clear expectation that any single surface brightness limit can
distinguish galaxies from intracluster light, we explore a range of
thresholds ($\mu_{limit}$ = 26 -- 27.5) for characterizing the diffuse
light in clusters.

For this exercise we first identified all stellar sources on each
cluster image, and masked them out using our large-scale PSF.  We defined
a stellar source as all objects from our earlier SExtractor catalogs
that had a star/galaxy classifier value greater than 0.8.  We then
masked each individual pixel that had a corresponding surface
brightness brighter than some limit  $\mu_{limit}$.  We used four surface
brightness limits, to bracket the respective depths of our 
cluster images ($\mu_{limit}$ = 26,
26.5, 27, and 27.5).  In order to avoid biasing our results to negative
flux, we also masked the corresponding negative pixels to the 
identical negative ADU value as the positive surface
brightness limit.  This ensured that the mean 
flux of an empty region of sky approached zero as the 
isophotal limit was reduced.  
An image of the core of Abell~1234 masked in this manner is 
shown in Figure~\ref{fig:masking}.

To determine the relative fraction of the ICL light compared to the
galaxy light, we then created an inverse mask to isolate and measure
galaxy light -- in this inverse mask, point sources and all pixels
fainter than $\mu_{limit}$ were masked instead.  We then summed the
un-masked pixels in each mask, and determined the flux ratio for the
cluster region, after correcting for the area lost to the ICL from the
cluster galaxies. The errors in
the flux ratio were determined by our flux error model, and are
completely dominated by the large-scale flat-fielding errors.  We note
that this method gives us a luminosity-weighted average ratio over the
entire cluster, and washes away any radial structure in the ICL. 

Table~5 gives the fractional flux in the ICL for each
value of $\mu_{limit}$.  We show the results for Table~5 
graphically in Figure~\ref{fig:fractions}.  
The corresponding luminosities and luminosity densities are 
given in Table~6.  We find that the fraction of ICL 
measured in this way varies from 28\%
to 5\% when we restrict our measurements to above $\mu_{5\sigma}$.
Our measurement of Abell~801 at $\mu_{limit} = 26.0$ is 16 $\pm$
4.7\%, in remarkable agreement with early measurements by Baum (1973),
who found a fraction of 16\% at the same isophotal cutoff using
photoelectric photometry with single-channel scanning of
the entire cluster.  The ICL luminosities of each cluster vary from
1 $\times 10^{11}$ L$_{\sun}$ to 0.3 $\times 10^{11}$ L$_{\sun}$,
roughly equal to that of a few Milky Way galaxies.  The
luminosity densities found are roughly equivalent to those found
from intracluster star studies seen from nearby clusters 
(Durrell \etal 2002; Feldmeier \etal 2004).  

There are a number of additional uncertainties to these results that
are difficult to quantify precisely.  First, since we masked all
unresolved sources in the cluster images, we might exclude
unresolved galaxies in each cluster.  This would tend to artificially
increase our measurement of the ICL fraction.  However, at the same
time, we have included all non-stellar objects in the frame as
``cluster'' galaxies.  Since there are a number of foreground and
background galaxies in each cluster frame, including them would 
tend to decrease the fractional ICL flux determined.  

From inspection of our non-stellar luminosity functions from the
SExtractor catalogs, our observations reach at least five
magnitudes below M$^{*}$ \citep{lf1976} 
for these clusters, making any unresolved
galaxies a small portion of the total cluster luminosity.  If we
assume that the luminosity function of our clusters are similar to
that measured in Virgo by Trentham \& Hodgkin (2002), we find that at
most 2\% of the cluster luminosity can be lost in this
way. {Similarly, from spectroscopic observations of cluster galaxies
at similar redshifts, \citet[][]{wilson1997} estimate the contamination
rate for bright galaxies to be no higher than 25\% over the entire
cluster}, and is likely to be significantly smaller in the cluster
core.

However, the largest significant uncertainty is likely to be the
choice of isophotal cutoff for the measurement.  Galaxies are not all
truncated at a fixed surface brightness limit, and an examination of
Figure~\ref{fig:masking} clearly shows what appears to be galaxy light
``leaking'' around the edges of the isophotal masks for brighter
galaxies.  To estimate the magnitude of this effect, we quantify for
a fiducial elliptical galaxy the fractional luminosity located outside
the different isophotal cutoffs we adopt.  Since most galaxies in the
cluster cores are early-type, this should give us a first-order
approximation to the amount of light we could be misidentifying as
ICL.  We use the extremely well-studied galaxy NGC 3379, which has
been found to follow a \citet{gdv1948} profile over ten magnitudes of
surface brightness.   
We take the surface photometry of Capaccioli
\etal (1990), assuming a (B--V) color of 0.96 (de Vaucouleurs \etal
1991), and applying the best fit \rquart model, we determine the
fraction of galaxy light outside of each isophote.  This fraction is
given in the row entitled ``NGC 3379'' in Table~5.

We find that up to 8\% of the total measured ICL could be attributed to the
extended nature of galaxies.  However, this is likely to be a firm
upper limit, and is strongly dependent on the exact radial extent of
galaxies in clusters.  Since galaxies in clusters are truncated by the
mean tidal field of the cluster \cite[\ie][]{mer1984,gnedin2003}, it
is likely that the amount of galaxy light outside a particular
$\mu_{limit}$ is less than calculated for the galaxy NGC~3379.  
NGC 3379 is contained within the Leo~I galaxy
group, making it unlikely that it has been severely tidally truncated. 
Since the actual tidal radius for each galaxy depends sensitively on the 
velocity dispersion of the cluster and the velocity dispersion and 
orbital properties of the galaxy, we adopt a small radius to 
place a limit on this effect.  We take the results of 
\citet{mer1984}, which give the smallest tidal radii of around
$\sim$ 20~kpc.  If we artificially truncate the NGC~3379 results to 
this radius, which corresponds to a surface brightness of 
\muv ~= 26.5, we obtain the results of the row entitled 
``NGC 3379 (trunc)'' in Table~5.  In this case, the extended 
galaxy light contributes substantially less to the measurement 
of the fractional ICL flux.

Given the large uncertainty in the above effects, we conclude that the
average fraction of ICL in these clusters is approximately 10\%, and
certainly no more than 20\% of the light seen in galaxies.  
This is consistent with studies of intracluster light in nearby clusters
\citep{1996arna,ipn1,ftv1998,durr2002,ipn3} but is less than the measured
amount for the Coma Cluster (50\%; Bernstein \etal 1995), and less
than found for clusters with large cD galaxies ($\sim$ 20--40\%; Paper~I;
Schombert 1988).

\subsection{The Spatial Distribution of the ICL}

To understand the basic properties of the spatial distribution of the
ICL in these clusters, we began by visually examining the data.  Using
the identical binning procedure used above, we binned each cluster
image into small bins, including the galaxy light, but excluding all
identified point sources.  By experimentation, we found that a bin
size of 11 $\times$ 11 pixels was the best compromise between spatial
resolution and signal-to-noise for this purpose.  These images are
plotted in Figure~\ref{fig:coloricl}.
In order to highlight structure at different surface brightness
levels, we have color-coded the images by surface brightness: black
shows bins with surface brightnesses from \muv ~= 20 to 24, red shows
\muv ~= 24 to 26, green shows \muv ~= 26 to 27, and blue shows \muv ~= 26
to 27.  Masked bins, or bins with surface brightnesses below \muv ~=
28, are left uncolored.  For Abell~1553 and Abell~1914, where the
large-scale flat fielding errors are significantly larger, we do not
color the \muv ~= 27 to 28  bins.  By adopting these limits, we
ensure that all bins that are colored in Figure~\ref{fig:coloricl}
are at least 2$\sigma$ above the sky
background.

There is a striking difference between the clusters in the
spatial distribution of their ICL.  In the case of Abell~801 and
Abell~1553, the ICL appears to closely follow the galaxy light, and in
Abell~801 at least, appears to have elliptical symmetry. {Generally,
the differing isophotal levels have similar geometric structure in
these clusters.}  In contrast, Abell~1234 and Abell~1914 show a
very different behavior.  In these clusters, the ICL follows the
galaxy light less well, and there are asymmetric features that are
clearly visible in the images.  The size of these features are
large, on order of a few arcminutes, corresponding to a linear size of
a few hundred kpc. In these clusters, the different isophotal
levels have markedly different morphologies.

To confirm and quantify these qualitative results, we 
determined the median spatial centroid of the bins in each cluster at
differing surface brightness limits corresponding to the levels shown
in Figure~\ref{fig:coloricl}. We note
that this centroid is an unweighted median of the bin positions at 
each isophotal level, and is not a luminosity-weighted median.
To avoid contamination from background
sources, we limited the centroid determination to the regions used
in measuring the total ICL in \S 6.1.  We determined the errors in 
our centroids by bootstrap resampling the data 50,000 times, and
measuring the standard deviation of these determinations.  We found
that in every case, the error of the median centroid 
was half of a bin size or less.  We also tested the median 
centroid on an artificial 
star constructed from our large-scale PSF (\S 5.1), and found that
the centroid moved less than a few pixels, over the range of surface
brightness of interest.  Since the values for the median are 
by construction quantized on a 11 $\times$ 11 
pixel grid, we took a conservative value for the error of the 
centroid as 11 pixels (3.3\arcsec).  The positions for each
of the centroids are given in Figure~\ref{fig:centroid}.  

We found that for both Abell~1234 and Abell~1914, there were
significant (4$\sigma$ and 6$\sigma$, respectively) displacements of
the centroid in both the north-south and east-west directions from the
high surface brightness centroids to the lower surface brightness
ones.  Specifically, Abell~1234's measured centroid moved $\approx
16$\arcsec~to the southeast when the surface brightness limit was
extended from \muv ~= 24 to \muv ~= 28.  This shift can be clearly be
seen as due to a low surface brightness feature (colored in blue in
Figure~\ref{fig:coloricl}) that is approximately perpendicular to the
main cluster axis (colored in red and green bins), and is displaced to
the east.  Abell~1914's centroid moved $\approx 28$\arcsec~northeast
as the surface brightness limit was increased to \muv ~= 26, and then
moved $\approx 20$\arcsec~ southwest, as the surface brightness limits
were increased to \muv ~= 27.  These shifts can can be attributed to
the large envelope of light surrounding the cluster galaxies that is
clearly offset from the galaxy distribution.  By contrast, Abell~801
and Abell~1234's centroids remained within 2$\sigma$ of the original
determination at all surface brightness levels.  There appears to be
no visible pattern to the offsets derived to these clusters that
correspond with other surface brightness features.

In conclusion, we find a range of properties for the spatial distribution
of the intracluster light.  Clearly, in some cases, the cluster light
closely follows the galaxies, but in other cases, there are noticeable
changes in both the centroid and the shape of the luminosity distribution.

\subsection{The Search For Tidal Features}

With the global properties of intracluster light established for each
cluster, we now search for any large-scale tidal debris.  We emphasize
that we are focusing on debris that is of significant size compared
to the extent of each cluster.  In all of the clusters, we find 
many morphologically unusual small sources, but without color or 
redshift information we cannot be confident that these sources belong to 
the cluster, and are not foreground or background objects.  
We therefore leave their study for the future.

For the purposes of this discussion, we define a tidal debris arc as
an extremely elongated (ellipticity $\geq 0.5$) discrete object that
can be detected visually, and a tidal plume as any broader diffuse
emission not fixed to any single galaxy.  Two of us (J.F., C.M.),
searched each cluster image independently, and noted any features from
a visual inspection that were detected in common.  We note that
searching for arc-like features will be strongly dependent on the
seeing, and so we expect fewer features detected in Abell~801 and
Abell~1553 than in the other clusters.  However, larger plume-shaped
features should still be visible in all clusters, since their spatial
scale is much larger than the seeing disk.

In both Abell~801 and Abell~1234, we found no large-scale discrete
tidal features that could be attributed to the intracluster light.  We
did detect a few smaller features around the central elliptical
galaxies in Abell~801 that resemble the low surface brightness
``shells'' seen by Malin \& Carter (1983), but due to the poor seeing
more study will be needed to confirm these objects.  In Abell~1234 we
find a number of small-scale diffuse objects, but none of these can
clearly be attributed to tidal features or gravitational lensing based
on our imaging alone. In both Abell~801 and Abell~1234, we do find
that the galaxies in general appear to be surrounded by a common
envelope of diffuse intracluster light.

In contrast, Abell~1553 has a bright plume-like feature extending from
the southeast of one of the central galaxies, which is plotted in
Figure~\ref{fig:a1553plume}.  It can also be clearly seen in
Figure~\ref{fig:coloricl} as the green region just above the masked
star in the center of the image.  From inspection of the original
image (Figure~\ref{fig:a1553plume}, left), there is a roughly
semi-circular plume (hereafter referred to as the bright plume) that
is $\approx 13 \arcsec$ in radius, and has a mean surface brightness
of \muv ~$\approx$ 25.  After inspection of the binned-up data, we
also found a fainter plume-like object (hereafter referred to as the
faint plume) that is along the path of the bright plume, but had a
wider opening angle, and a much fainter mean surface brightness (\muv
~$\approx 26.2$ ).  We plot this fainter plume in the right panel of
Figure~\ref{fig:a1553plume}, where we denote the area of the plume as
the spatial bins with white central pixels.  There is an abrupt change
in surface brightness marking the transition between these two
features, making it difficult to determine whether they comprise a
single object, or are distinct structures.  We measured the apparent
magnitudes of both plumes, and found a total magnitude of $m_{v} \sim
19.0$ for the bright plume and $m_{v} \sim 18.8$ for the faint plumes,
corresponding to an absolute magnitude of $V \sim -20.3 $ and $-20.5$.
These are each approximately a Milky Way's worth of luminosity, spread
out over an areas of approximately 2000 kpc$^{2}$ and 5000 kpc$^{2}$,
respectively.  Curiously, at the location of the faint plume the
distribution of cluster galaxies appears to turn abruptly, from
running northwest to southeast, to extending due east.  It is unclear
whether this is a coincidence or has some relation to the tidal plume
structure.

In Abell~1914, there is a wealth of diffuse structure that is
immediately apparent to the eye.  We detect five large-scale features:
two arc-like features, one to the west of the cluster core, and one in
the geometric center of the cluster, and three tidal plumes each
leading from the triangle formed from the brightest galaxies in the
cluster, which we denote as the northwest plume, the southwest plume,
and the eastern plume, respectively.  They are all displayed in
Figure~\ref{fig:a1914global}, and are discussed in detail below.

The western arc is displayed in Figure~\ref{fig:westarc}.  The arc is
approximately 1\arcmin ~long ($\approx$ 160 kpc), and has a maximum
width of 12\arcsec ~($\approx$ 30 kpc), though in portions the arc
becomes as narrow as 2.2\arcsec ~($\approx$ 6 kpc).  The arc appears
to have a bulge or bifurcation approximately halfway along its length.
The maximum surface brightness of the arc is \muv $\approx$ 25.4, and
the median surface brightness is \muv $\approx$ 26.1.  The properties
of this arc candidate in physical size and in surface brightness are
very similar to the tidal debris arcs seen in the nearby Coma and
Centaurus clusters \citep[][; see Paper~I, Table 5 for a
summary]{tren1998,gregg1998,cr2000}.  Although \citet{dlens2002} note
the presence of several gravitational arc candidates in their images
of Abell~1914, we believe that it is unlikely that this particular arc
is due to gravitational lensing.  Specifically, the arc is radially
resolved, widens significantly along its path, and is not tangential
to the mass distribution.  The cause of the bulge or bifurcation
midway through the arc candidate is unknown: it may be due to a
gravitational disturbance from a passing galaxy.

The central arc is 5.4\arcsec ~long and 1.7\arcsec ~wide.  It is
visually unremarkable, and appears almost rectangular.  It has a
larger mean surface brightness \muv $\approx$ 25, and a peak
surface brightness of \muv $\approx$ 24.6.  This arc may well
be due to gravitational lensing: it is short, tangential 
to a portion of the mass distribution, and is radially narrower than 
the western arc.  It appears to lie within a common envelope of 
the nearby galaxies.  

The southwest plume is extremely large, more akin to a extended
envelope, and can be seen most clearly in Figure~\ref{fig:coloricl}
as the extremely large asymmetry colored in green to the upper right
of the cluster center.  This structure lies along one of the cluster
axes, and a complex of several luminous galaxies lies within it.  It
has an enormous extent, at least 98\arcsec ~along the axis of the
cluster and 112\arcsec ~wide, corresponding to an approximate physical
size of 270 by 300 kpc. It has a mean surface brightness of \muv
$\approx$ 26.7.  However, due to its spatial proximity to the edge of
the large-scale flat fielding pattern, its irregular structure, and
its low surface brightness, we do not attempt to determine a total
magnitude for this feature, though it should be quite large.

The northwest plume can be seen as a narrow, roughly triangular
plume extending away from one of the central elliptical galaxies.
It has a mean surface brightness of \muv $\approx$ 26, and has
a width of $\approx 11\arcsec$, and a height of $\approx 22\arcsec$,
corresponding to physical dimensions of $\approx$ 30 and 60 kpc, 
respectively.  The total magnitude of this feature is $m_{v} \approx
20.7$, corresponding to an absolute magnitude comparable to a small
spiral galaxy.  

Of all the tidal features seen however, the most striking is the large
plume to the east of the cluster core, which is displayed in
Figure~\ref{fig:eastplume}.  The plume is located at the end of a
linear structure of six luminous elliptical/S0 galaxies that show
signs of tidal interaction, and lie within a common surface brightness
envelope.  The eastern plume has a size of 52\arcsec ~in the east-west
direction, and 39\arcsec ~in the north-south direction.  The mean
surface brightness of this feature is \muv $\approx$ 26.4.  The total
magnitude from this plume is enormous, V $\sim -21.3$, comparable to
the entire optical luminosity of M31. The plume appears sharp-edged --
the surface brightness drops by over a factor of two (down to the
background level) over 3.4\arcsec\ (9 kpc) at the southern edge.  Such
a sharp edge strongly implies a recent tidal origin for this feature,
since if it had been torn out early in the cluster's history, or in
the outskirts of the cluster, any sharp features would been
dynamically heated and mixed away by the tidal field of the cluster
and its member galaxies.

The most remarkable feature of this plume, however, is that it lies on
top of a feature seen through gravitational lensing.  Specifically,
the eastern plume can be clearly seen in the weak lensing map of
\citet{dlens2002}, a property not seen for the other plumes in
Abell~1914.  This implies that a large fraction of matter in addition
to the optically luminous matter must be associated with this
particular plume, since the intracluster light is a minuscule part of
the cluster's total mass.

Because this plume does appear to trace mass as inferred from the
lensing maps, we might also expect to observe it in X-ray emission.
We have compared an 8.7ks {\sl ROSAT} PSPC archival image (PI:
B\"{o}hringer) of Abell~1914 to our optical image, binned to the same
spatial resolution (15 arcseconds) as the PSPC image.  We find, as
Jones \etal (2001) did, that the X-ray emission from Abell~1914 shows
significant substructure in the core and there appears to be a small
extension in flux to the east, which might correspond to the linear
structure of the six elliptical/S0 galaxies mentioned above.  While we
see no significant emission associated with the plume itself, the
signal-to-noise of the X-ray data is relatively low, making it
difficult to set meaningful limits.  Further observations of
Abell~1914 with higher resolution X-ray telescopes, such as {\sl
Chandra} may prove more revealing.

In conclusion, we find that the amount of clearly identified tidal
debris at our observational limits varies significantly from cluster
to cluster.  However, some clusters have strong tidal activity
that ranges over physical scales of hundreds of kiloparsecs, and in
at least one case, follows an asymmetric dark matter distribution, 
perhaps due to the stripping of dark matter from galaxy halos.  

\section{Comparing to Simulations}

To place our observational study of the ICL in a more dynamical
context, we compare our deep imaging to simulations of cluster
collapse in a cosmological setting. Most previous simulations of tidal
stripping and ICL formation have been based on simulations of
individual galaxies moving through an established cluster potential
well (\eg ~Richstone \& Malumuth 1983; Miller 1983; Merritt 1983, 1984). 
However, the dynamics of cluster formation and tidal
stripping are much more complex -- clusters grow hierarchically, with
small groups of galaxies accreted as the cluster grows larger and
larger. Within these small groups, the tidal interactions can be
slower and more dramatic, liberating a substantial amount of
intracluster starlight (\eg ~Mihos 2003). More recent simulations have
studied the evolution of galaxies in a dynamically evolving cluster,
giving a more realistic view of the forming ICL.

To visualize the formation of diffuse light in galaxy clusters, we use
N-body simulations of galaxy clusters from Dubinski \etal (2003).
Full details of the simulation technique can be found in Dubinski
(1998) and Dubinski \etal (2003); we summarize the salient points
here. The simulations were created by first evolving a cosmological
$\Omega_M=1$ dark matter simulation, identifying bound clusters ($M
\sim 10^{14}M_{\sun}$, $\sigma_v ~ 300-600$ km/s) at $z=0$, and
replacing the 300 most massive halos at $z=3$ destined to lie within
the $z=0$ cluster with composite disk/bulge/halo galaxy models. Upon
replacement, the galaxy models are scaled in mass and size such that
they retain a constant surface density and follow the Tully-Fisher
relationship (see Dubinski 1998 and Dubinski \etal 2003 for
details). A total of 3,600,000 particles comprise the luminous
portions of the galaxies; these particles were used to construct
artificial images of the clusters. While the use of an $\Omega_M=1$
simulation may seem questionable in the light of modern $\Lambda$
cosmologies, it should have only modest impact on our main results.
In the evolution of rich clusters, cosmology contributes largely
through the evolving accretion rate onto the cluster, so that these
simulations will have proportionally larger late-time accretion than
in $\Lambda$ cosmologies. We comment further on this effect below.

We create artificial images of these simulations by first spatially
binning the particles to a comparable physical scale as our CCD images
-- in this case, one pixel corresponds to 800 pc. We assign each pixel
a luminosity based on a V-band mass-to-light ratio of 5, typical of an
old stellar population.  We then smooth the image using a locally
varying smoothing kernel which depends on both the mass density within
the pixels and the ``filling factor'' of neighboring pixels. The
intent of this smoothing is to reduce the graininess inherent in
$N$-body simulations, but we run the risk of over-smoothing the data
in regions of very low density, artificially creating diffuse extended
features. To prevent this, we limit the smoothing kernel in low
density regions to a maximum $\sigma=8$ kpc. After this step, we rebin
in 11x11 pixels and calculate a robust mean surface brightness in an
identical fashion to that done for the real observational data.

The simulated cluster images are shown in Figure~\ref{fig:simimages}, 
color coded by
surface brightness in the same fashion as the real observed clusters
in Figure~\ref{fig:coloricl}.
Each frame is approximately 2.3 Mpc on a side. The
greyscale intensities show surface brightnesses below \muv ~= 28 
, where the structure of the images begins to become
mottled by discreteness noise in the simulation. The faintest
structures seen in these images have \muv ~$\sim$ 30,
well below our observational detection limit.  The top four panels show the
evolution of an individual cluster over time. We caution that these
images are not meant to simulate the appearance of high redshift
clusters -- we do not include $(1+z)^4$ dimming, K-corrections, or an
evolving mass-to-light ratio for the stellar populations. Instead, we
are simply looking at how dynamical evolution and tidal stripping
build up the diffuse ICL, and redshift is used here only as a
dynamical clock for the cluster.  The bottom panels show four
different simulated clusters at $z=0$, illustrating the wide variety
of morphological features shown by evolved clusters.

Comparing our observational data to the simulations, it is clear we
are just glimpsing the tip of the iceberg in terms of the spatial
structure of the ICL. At \muv ~$\sim 26$, the isophotes are fairly
regular; slight asymmetries show up at \muv ~$\sim 27$ -- compare the
irregular envelope of cluster C6 with similar features in the cD
cluster Abell~1413 (Paper~I). At \muv ~$\sim 27$ we also see 
the kind of ``common
envelope'' features seen in some of our real systems, most notably
Abell~1553.  At the limit of our current detections, \muv ~$\sim 28$,
more asymmetries show up in the evolved clusters, such as the plumes to
the lower right in C2 and C4. However, the most interesting dynamical
features -- large scale plumes and streams -- show up at even lower
surface brightnesses, \muv ~$\sim 30$. To probe these structures, much
deeper imaging is needed; here, issues of flat fielding, scattered
light, and sky variability will become severe.

Figure~\ref{fig:simimages} also shows how the morphology of 
the ICL changes with dynamical 
time. Early in the cluster's history, as the bulk of the cluster is
being assembled, significant tidal stripping results in plumes and
arcs at relatively high surface brightness, \muv ~$< 26$. These features
can be quite thin and sharp, and such features are likely a sign of a
major, recent stripping event in the cluster. The relatively high
surface brightness arcs seen in several nearby clusters (Trentham \&
Mobasher 1998; Gregg \& West 1998; 
Calcaneo-Roldan \etal 2000) are likely examples of
recent stripping.  As the cluster evolves, many of these features
``mix away,'' creating a smoother ICL at these surface brightnesses in
the more evolved clusters. In this context it is again interesting to
note that we see more substructure in our non-cD clusters presented
here than in the cD clusters of Paper I. Inasmuch as the presence of a
cD galaxy indicates an evolved cluster, this difference in ICL
properties between cD and non-cD clusters is as expected from these
evolutionary models.

We can quantify these isophotal asymmetries in a similar fashion to
that done for the observational data -- through a measurement of the
centroid shift as a function of surface brightness. As with the
observational data, we mask regions well outside the main body of the
cluster and then calculate the isophotal centroids. In the 9 simulated
clusters, we found spatial offsets between the \muv ~= 24 and \muv ~= 26 
isophote which ranged from 0--35 kpc, and from 10--65 kpc for the
offset between \muv ~= 24 and \muv ~= 28. The cluster-to-cluster variation
is large, with some very smooth clusters (like cluster C4 in 
Figure~\ref{fig:simimages})
showing little or no offset at all. These spatial offsets are similar
to those seen in our Abell clusters -- the large offsets are
comparable to those observed in Abell 1234 and Abell 1914, while
several of our clusters show no significant offsets at all.

Finally, we can look at the fraction of material in the simulated
images as a function of limiting surface brightness. This measurement
is perhaps the most uncertain, since it depends both on the adopted
M/L value for the stars {\it and} the assumption that this value is
fixed across the cluster. Nonetheless, we can get a simple estimate of
the amount of material in the ICL for comparison to our
observationally-derived numbers in Table~5.  Figure~\ref{fig:simfrac}
shows the fractional ICL as a function of limiting surface brightness
for the nine evolved cluster simulations of Dubinski \etal (2003). At
high surface brightness, \muv ~$\sim$ 26, 8--18\% of the cluster
luminosity is in the diffuse component; this number drops to $\sim$
3--6\% at \muv ~= 28.  At low surface brightness, our observed ICL
fraction is lower than the simulations, likely because we are simply
nearing our detection limit.  At higher surface brightness, we
actually detect more ICL than simulated -- a number of effects could
be causing this. First, background contamination in the observational
data would increase the amount of flux at low surface brightness.
Secondly, and more importantly, the simulation data employs only
luminous disk galaxies.  A population of low surface brightness dwarf
galaxies, or even the extended low surface brightness envelopes of
elliptical galaxies, would raise the diffuse light content of the
simulated galaxies significantly.  Figure~\ref{fig:simfrac} shows the
evolution of the amount of diffuse light in the evolving
cluster. Early in the cluster's history, as the bulk of the cluster is
assembled, the diffuse light content increases quickly. At later times
the ICL fraction oscillates, due largely to the accretion of infalling
groups. As groups fall in to the cluster environment, they first
reduce the fractional ICL by raising the high surface brightness
content of the cluster. As they then orbit through the cluster and are
tidally stripped, the diffuse light content of the cluster rises
again. The magnitude of these fluctuations may well be a feature of
the $\Omega_M = 1$ cosmology, with its significant late accretion --
clusters in a low $\Omega_M$ universe may show this effect at a much
lower level.

Any more quantitative analysis needs more sophisticated models
employing a modern $\Lambda$ cosmology and mix of progenitor types --
models which are currently being developed (\eg ~Napolitano \etal 2003;
Dubinski \etal 2003; Mihos et~al., in
preparation). Nonetheless, even the simulated images presented here
reveal striking similarities with the observational data -- to the
depths that we currently probe, the detected morphology shows many of
the structural features (common envelopes, faint arcs and plumes, and
isophotal centroid shifts) evidenced in the dynamical models. However,
these simulations also point toward the wealth of information
contained in the detailed morphology and kinematics of the ICL lying
just out of reach of current observational data. Deeper imaging, along
with followup spectroscopy of PNe in nearby clusters 
to determine the kinematics of diffuse light, is a promising avenue 
toward unraveling the structure of the ICL.

\section{Summary}

We have surveyed four Abell type II-III (non-cD) galaxy clusters, and
find a significant amount of intracluster light in all of them.  The
amount of intracluster starlight is intermediate, approximately
10--20\% of the galaxy light.  This may reflect the dynamical
adolescence of these galaxy clusters: as the clusters continue to
evolve the mean intracluster fraction should increase.  However, it is
likely that cluster mass (or richness) also plays a role: both our own
measurements of the rich cluster Abell~1413 (Paper~I), and other
studies of richer clusters such as Coma (Bernstein \etal 1995) find
significantly higher intracluster star fractions than the clusters
studied in this paper.  Clearly, more observations of clusters over a
range of cluster richnesses will be important to separate the two
related effects.

We have searched for signs of asymmetries and tidal debris in the
intracluster light, and we find multiple examples of these features,
though in some clusters we found signs of a regular symmetric
structure in the intracluster light and few tidal features.  The
number of tidal features in Abell~1914 is clearly unusual, in both
number and surface brightness.  It is worth reiterating here that
Abell~1914 has a radio halo\citep{gio1999,kemp2001} and is believed to
be undergoing a cluster merger (Jones \etal 2001).  Therefore, the
wealth of high surface brightness tidal features in the ICL in
Abell~1914 is likely to be due to the ongoing cluster merger.  We may
be observing the cluster in {\it flagrante delicto}: had we observed
the cluster a few crossing times later, the evidence of such a strong
merger would be much less apparent.

Comparing our observational results to modern numerical simulations of
galaxy clusters, we find good overall qualitative agreement in both
the amount and distribution of the intracluster light, although the
simulations show that significant ICL substructure may exist well
below out current detection levels.  From the results found here, we
suggest that intracluster light may act as a dynamical ``clock'' of
galaxy clusters, one that is complementary to other studies of galaxy
clusters such as X-ray substructure, weak lensing, and galaxy radial
velocities.  However, the comparison of ICL properties and models of
structure formation and evolution is still in its infancy.  A detailed
comparison of the two will require much deeper ICL imaging and
measurements of the kinematics of the ICL (though planetary nebulae
velocities, for example), as well as through numerical simulations
that have larger dynamical range, and better descriptions of 
galaxy formation processes.  Such studies are now underway.

\acknowledgments

We thank the KPNO staff for their assistance with the observations.
We especially thank William Baum for his communication regarding the
early study of Abell~801.  We also thank Cameron McBride for a large,
yet oddly undefinable, range of technical and scientific support.
This work is supported by NSF through grants AST-9876143 (JCM),
AST-9624542 (HLM), and AST-0302030 (JF).  Funding was also given by
the Research Corporation's Cottrell Scholarships (JCM and HLM).  
We also thank an anonymous referee for suggestions that improved
the presentation of this data.  This
research has made use of the USNOFS Image and Catalogue Archive
operated by the United States Naval Observatory, Flagstaff Station
(http://www.nofs.navy.mil/data/fchpix/).  This research has 
made use of data obtained through the High Energy 
Astrophysics Science Archive Research Center Online Service, 
provided by the NASA/Goddard Space Flight Center.

\pagebreak
\appendix
\section{The Error Model}

It is necessary to have accurate error estimates of our surface 
photometry in order to perform the model fitting.  
Unlike earlier photographic work, deep CCD surface photometry
allows us to quantify measurement errors.  Measurement errors arise
from CCD behaviors such as readout noise and flat-fielding, as well from
sky noise.  Each error contribution will be addressed below.

\section{Readout Noise}
The readout noise per exposure is 1.1 ADU.  By combining $N_{G}$ images
with a median, we are able to reduce the effective read noise to
\begin{equation}
R_{eff} = 1.1~ADU~\frac{1.22}{\sqrt{N_{G}}} 
\end{equation}

The factor of 1.22 is due to the lower efficiency of a median over
a mean (see Morrison, Boroson, \& Harding 1994, Section 3.2.1).

\section{Photon Noise}
For $C$ ADU in a given pixel, the photon noise is (C/$g$)$^{1/2}$, where 
$g$ is the gain.  Combining $N_{G}$ images respectively using a 
median reduces the photon noise to
\begin{equation}
\sigma_{Poisson} = \frac{1.22}{\sqrt{N_{G}}} \frac{\sqrt{C}}{\sqrt{g}}
\end{equation}

\section{Linearity Errors}
As mentioned in \S5.2 above, the T2KA chip has a known non-linearity.
We have corrected for this effect, but the parameters used for the
correction do not have infinite precision, and so our correction 
has errors.  The error in flux can be derived as follows:

\begin{equation}
\sigma_{linearity}^{2} = \sigma_{c1}^{2} C_{sky}^{2} + \sigma_{c2}^{2} 
\frac{C_{sky}^{4}}
{(32767)^{2}} + \sigma_{c3}^{2} \frac{C_{sky}^{6}} {(32767)^{4}}
\end{equation}
where $C_{sky}$ is the sky-subtracted flux.  Since this correction is small,
we apply it only to the flux, and not to any other calibration image. 
\section{Flat-Fielding Errors}
In principle, the only limit to the precision of the combined flat-field
images is the photon noise in the individual flat-field images.  This
small-scale variation is
\begin{equation}
\sigma_{sff} = \frac{\sqrt{C_{s}}}{\sqrt{g}} 
\frac{1.22}{\sqrt{N_{f}}} \frac{1.22}{\sqrt{N_{G}}}
\end{equation}
where $C_{s}$ is the number of counts in the final, combined master sky
flat image, g is the gain, $N_{f}$ is the number of individual sky flats
used to make the master sky flat, and N$_{G}$ is the number of individual
galaxy images used to make the final galaxy image.  

In practice, the small-scale flat-fielding errors are not the only 
flat-fielding error we have.  There are also large-scale variations which
arise from the variation of the sky brightness across the image, 
instrumental effects such as flexure, and from the wings of bright
stars and galaxies that were not completely removed by combining
the individual sky flats.  Normally, to measure this effect, we prefer
to divide our sky flats into two sub-samples, create two sky flat
images from those sub-samples, and then find the standard deviation 
of the ratio of the two created flats.  However, we have few 
sky images, and dividing them up into two ten image sub-samples would be
too noisy for a realistic measurement. 

Instead, we masked each image, and constructed a histogram of sky
values (\S 5.2), and determined the large-scale flat-fielding
error from those histograms.  Generally, this large-scale flat-fielding
error is the dominant source of error at low surface brightnesses.

\section{Surface Brightness Fluctuations}
For ultra-deep surface brightness observations of nearby galaxies,
a major source of error arises from intrinsic surface brightness
variations \citep{ts1988}.  For our distant galaxy clusters 
(see eq 10 of \citet{ts1988}), such an effect is completely 
negligible compared to our other errors.

\pagebreak

\pagebreak
\begin{deluxetable}{lllllcc}
\tablewidth{0pt}
\tablenum{1}
\tabletypesize{\footnotesize} 
\tablecaption{Target Information}
\tablehead{
\colhead{Cluster} 
& \colhead{RA\tablenotemark{a}} 
& \colhead{Dec\tablenotemark{a}}
& \colhead{Richness\tablenotemark{a}}
& \colhead{Redshift\tablenotemark{b}}
& \colhead{Bautz-Morgan type\tablenotemark{c}}
& \colhead{Rood-Sastry type\tablenotemark{d}}
}
\startdata
Abell~801  & 09h 28m 01.4s & +20d 33m 54s & 2 (C = 81)& 0.1918 & II-III & B (b)\\
Abell~1234 & 11h 22m 26.2s & +21d 23m 32s & 2 (C = 88)& 0.1663 & III & L (6)\\
Abell~1553 & 12h 30m 50.1s & +10d 34m 26s & 2 (C = 100)& 0.1652 & III & L (4)\\
Abell~1914 & 14h 26m 03.0s & +37d 49m 32s & 2 (C = 105)& 0.1712 & II: & L (6)\\
\enddata
\tablenotetext{a}{Coordinates are cluster
centers, C is the background-corrected count of cluster members 
between m$_{3}$ and m$_{3}$ +2, data taken from \citet{aco}.}
\tablenotetext{b}{Data taken from \citet{sr1999}}
\tablenotetext{c}{Data taken from \citet{leir1977}}
\tablenotetext{d}{Data taken from \citet{sr1987}}
\end{deluxetable}

\begin{deluxetable}{lcccccccc}
\tablewidth{0pt}
\tablenum{2}
\tabletypesize{\scriptsize} 
\tablecaption{Observing Log \& Adopted Target information}
\tablehead{
\colhead{Target} 
& \colhead{N$_{\rm exposures}$ }
& \colhead{Median sky}
& \colhead{Median seeing}
& \colhead{Average HA}
& \colhead{Image size\tablenotemark{b}}
& \colhead{Angular size\tablenotemark{c}}
& \colhead{Image size}
& \colhead{DM\tablenotemark{c}\tablenotemark{d}}\\
\colhead{}
& \colhead{}
& \colhead{(\muv)}
& \colhead{(arcseconds)}
& \colhead{(minutes)}
& \colhead{(arcminutes)}
& \colhead{Distance (Mpc)}
& \colhead{(Mpc)}
& \colhead{(mag)}
}
\startdata
Abell~801  &  6 & 21.2 & 2.6 & -38 $\pm$ 29 & 10.3 $\times$ 9.9  & 614.6 
& 1.8 $\times$ 1.8 & 39.7 \\ 
Abell~1234 & 14 & 21.3 & 1.1 & -80 $\pm$ 80 & 8.2 $\times$ 9.2  & 548.0
& 1.3 $\times$ 1.5 & 39.4 \\
Abell~1553 & 20 & 21.1 & 2.0 & 16 $\pm$ 70 & 9.8 $\times$ 10.1 & 545.0 
& 1.6 $\times$ 1.6 & 39.3 \\
Abell~1914 & 11 & 21.6 & 1.0 & 07 $\pm$ 70 & 10.0 $\times$ 9.9 & 561.6
& 1.6 $\times$ 1.6 & 39.4 \\
Blank Sky  & 49 & 21.3 & --   & -28 $\pm$ 96 & -- & -- & -- \\
\enddata
\tablenotetext{a}{The error bars show the spread in the hour angle distribution
for the source's exposures, not the error in the hour angle 
\hspace{\fill} \linebreak (which is negligible).}
\tablenotetext{b}{Image dimensions are E-W by N-S, and show the final trimmed
sizes of the images.}
\tablenotetext{c}{Assuming our adopted cosmology ($H_0=75$ km/s/Mpc, 
$\Omega_{\mbox{m}} = 0.3$, $\Omega_{\Lambda} = 0.7$)}
\tablenotetext{d}{Distance Modulus, assuming our adopted cosmology}

\end{deluxetable}

\begin{deluxetable}{lcccccc}
\tablewidth{0pt}
\tablenum{3}
\tabletypesize{\footnotesize} 
\tablecaption{Adopted Large-scale Flat-field errors and Observational
Limits}
\tablehead{
\colhead{Target} 
& \colhead{Adopted error} 
& \colhead{Notes} 
& \colhead{Percent Error} 
& \colhead{Total Bin error} 
& \colhead{$\mu_{1\sigma}$} 
& \colhead{$\mu_{5\sigma}$} 
\\
& \colhead{(ADU)}
& \colhead{}
}
\startdata
Abell~801  & 0.6 & standard deviation & 0.07 & 0.91 & 28.58 & 26.83\\
Abell~1234 & 0.6 & standard deviation & 0.08 & 0.62 & 29.00 & 27.25\\
Abell~1553 & 1.5 & visual inspection & 0.17  & 1.53 & 28.01 & 26.26\\
Abell~1914 & 2.0 & visual inspection & 0.35  & 2.02 & 27.71 & 25.96\\
\enddata
\end{deluxetable}

\begin{deluxetable}{lcc}
\tablewidth{0pt}
\tablenum{4}
\tabletypesize{\footnotesize} 
\tablecaption{Comparison of Surface Photometry for 1119+216}
\label{table:compare}
\tablehead{
\colhead{Parameter} 
& \colhead{Ledlow \& Owen (1995)} 
& \colhead{Our results} 
}
\startdata
r$_{24.5}$\tablenotemark{a} & 27.99 kpc & 29.0 $\pm$ 1.6 kpc \\
ellipticity (r$_{24.5}$) & 0.298 & 0.281 $\pm$ 0.029 \\
PA (r$_{24.5}$) & 85$\degr$ & 79.7$\degr$ $\pm$ 3.2$\degr$ \\
\enddata
\tablenotetext{a}{Radius given is (ab)$^{1/2}$}
\end{deluxetable}

\begin{deluxetable}{lcccc}
\tablewidth{0pt}
\tablenum{5}
\tablecaption{Isophotal ICL Fractions}
\tablehead{
\colhead{Target} 
& \colhead{$\mu_{limit}$ = 26.0} 
& \colhead{$\mu_{limit}$ = 26.5} 
& \colhead{$\mu_{limit}$ = 27.0} 
& \colhead{$\mu_{limit}$ = 27.5} 
}
\startdata
Abell~801 & $16 \pm 4.7$\% & $6.9 \pm 4.2$\% & 
$2.7 \pm 4.0$\%\tablenotemark{a}  & $1.2\pm 3.9$\%\tablenotemark{a}   \\ 
Abell~1234 & $17 \pm 4.4$\% & $10 \pm 4.1$\% & $4.9 \pm 3.9$\% & 
$2.1 \pm 3.7$\%\tablenotemark{a}  \\
Abell~1553 & $21 \pm 16$\% & $12 \pm 15$\%\tablenotemark{a}  
& $5.3 \pm 14$\%\tablenotemark{a}  & $2.2 \pm 13$\%\tablenotemark{a}  \\
Abell~1914 & $28 \pm 16$\% & $15 \pm 14$\%\tablenotemark{a}  & 
$6.8 \pm 13$\%\tablenotemark{a}  &$2.9 \pm 13$\%\tablenotemark{a}  \\
\hline
NGC 3379 & 7.8\% & 5.0\% & 3.9\% & 2.0\% \\
NGC 3379 (trunc) & 2.6\% & 0\% & 0\% & 0\% \\
\enddata
\tablenotetext{a}{This measurement is below $\mu_{5\sigma}$
for this cluster}
\end{deluxetable}

\begin{deluxetable}{lcccc}
\tablewidth{0pt}
\tablenum{6}
\tablecaption{ICL Luminosity and Luminosity Densities}
\tablehead{
\colhead{Target} 
& \colhead{$\mu_{limit}$ = 26.0} 
& \colhead{$\mu_{limit}$ = 26.5} 
& \colhead{$\mu_{limit}$ = 27.0} 
& \colhead{$\mu_{limit}$ = 27.5} 
}
\startdata
Luminosity ($10^{11} $L$_{\sun}$): \tablenotemark{a}& & & & \\
Abell~801  & $1 \pm 0.3$ & $0.8 \pm 0.5$  & $0.4 \pm 0.6$ & $0.3 \pm 1.0$ \\ 
Abell~1234 & $3 \pm 0.8$ & $2 \pm 0.8$ & $1 \pm 0.8$ & $0.8 \pm 1.4$ \\
Abell~1553 & $3 \pm 2$ & $2 \pm 3$ & $1 \pm 3$ & $0.7 \pm 4$\\
Abell~1914 & $8 \pm 5$ & $6 \pm 5$ & $4 \pm 8$ & $3 \pm 13$\\
\hline
Luminosity Density ($10^{5} $L$_{\sun} $kpc$^{-2}$):
\tablenotemark{a} & & & & \\
Abell~801  & $10 \pm 3$& $6 \pm 4$& $4 \pm 6$& $3 \pm 10$\\ 
Abell~1234 & $8 \pm 2$ & $6 \pm 2$& $4 \pm 3$& $3 \pm 5$\\
Abell~1553 & $8 \pm 6$ & $5 \pm 6$& $4 \pm 10$& $2 \pm 12$\\
Abell~1914 & $13 \pm 8$ & $10 \pm 9$ & $7 \pm 13$& $5 \pm 22$ \\
\enddata
\tablenotetext{a}{Assuming the adopted 
DM, but with no K-corrections.  For $V$ magnitudes, the K-corrections
are on order of 0.4 magnitudes at these redshifts (Coleman, Wu, \& Weedman 1980)}
\end{deluxetable}

\clearpage
\pagebreak
\begin{figure}
\figurenum{1}
\label{fig:clusters}
\caption{Our final, median-combined images for all of the observed 
clusters.  The top left cluster is Abell~801, top right is Abell~1234,
bottom left is Abell~1553, and the bottom right is Abell~1914.  
North is at the left of each image, and east is at the bottom.  
The image sizes are given in Table~2.}  
\end{figure}

\pagebreak

\begin{figure}
\figurenum{2}
\label{fig:psf}
\plotone{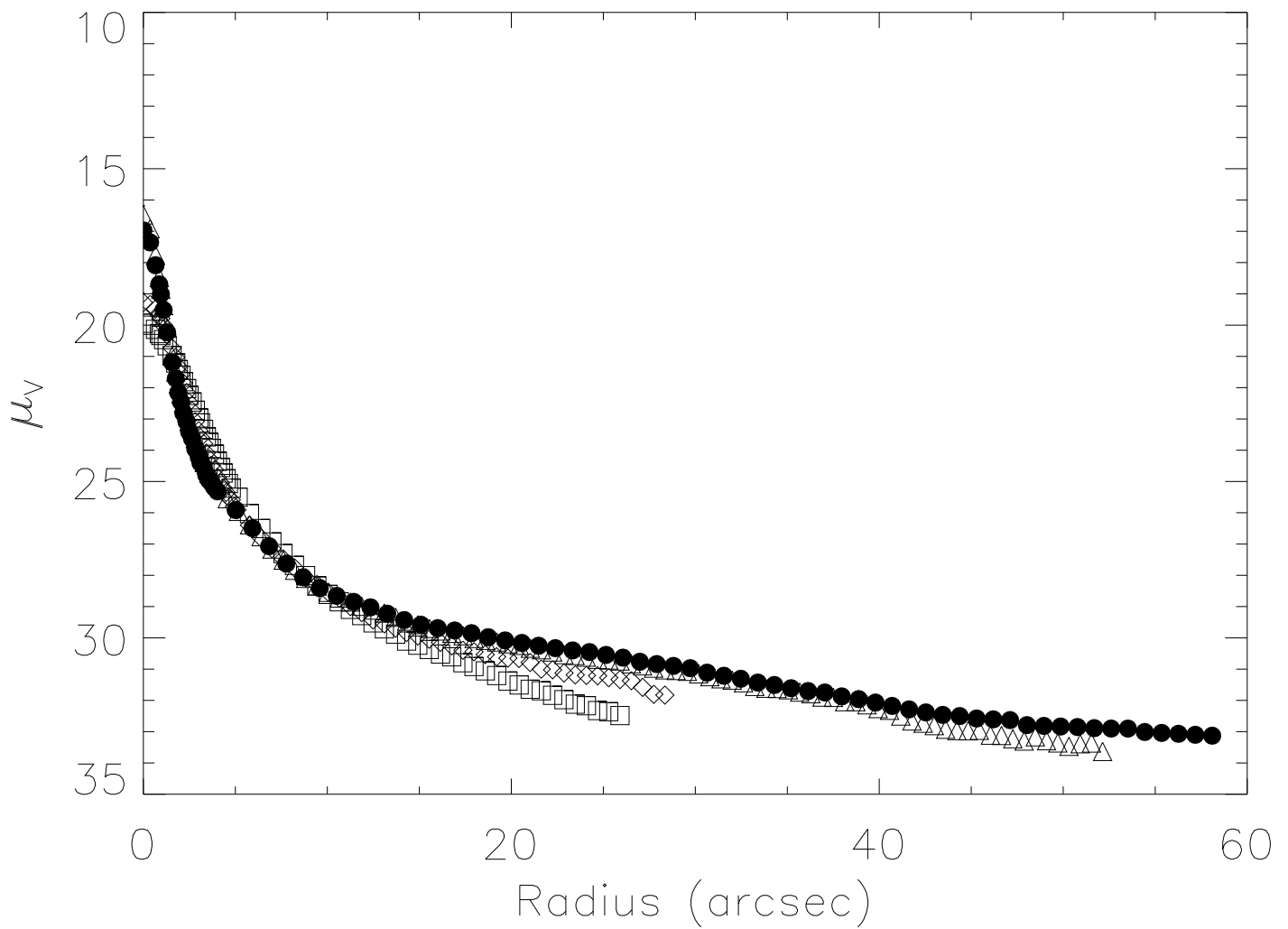}
\caption{The surface brightness profile of a saturated star on 
each of our cluster images, averaged azimuthally (see the text on how
these profiles were constructed).  For comparison purposes, 
a constant has been added to each profile to make them equivalent 
at R=10 arcseconds.  The filled circles denote the surface 
brightness profile from Abell~1234, the open diamonds 
Abell~1553, the open triangles from Abell~1914, and the open 
squares are the profile from Abell~801.  At small radii, the
profiles vary significantly due to seeing, but there is
excellent agreement at intermediate radii.  Since each saturated star
has a different apparent magnitude, the radius at where the profile
diverges and blends into the background noise varies.}
\end{figure}

\pagebreak
\begin{figure}
\figurenum{3}
\label{fig:hist}
\plotone{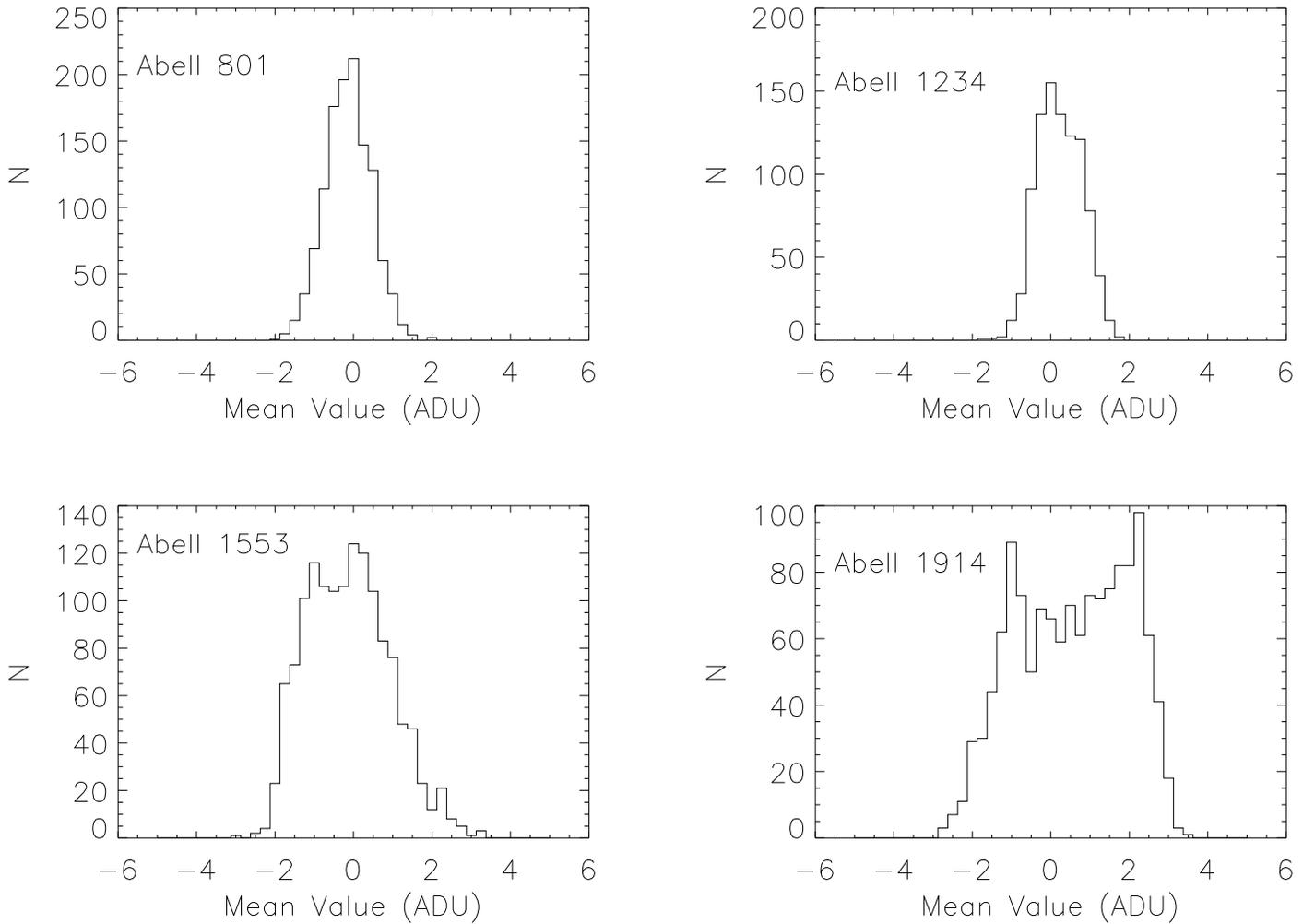}
\caption{The histogram of sky values for each cluster, binned up into 
0.25 ADU intervals.  See the text for the description of how this 
histogram was created. Ideally, the sky values should all equal zero, 
but due to large-scale flat-fielding errors, 
and the faint unmasked wings of stars and galaxies, there is usually a 
dispersion about zero in observed data.  Abell~801 
and Abell~1234 have well-behaved sky
histograms, with a well-defined mode, and whose probability distributions are
approximately Gaussian (see Paper~I, and Fry \etal (1999) for other 
such histograms), but Abell~1553 and Abell~1914 have broader sky 
histograms and are multi-modal due to large-scale flat fielding errors.  
See the text for further discussion.}
\end{figure}

\pagebreak
\begin{figure}
\figurenum{4}
\label{fig:flaterror}
\caption{A plot of sky values in Abell~1553 (top), and Abell~1914(bottom),
plotted as a function of row number.  The row number is equivalent to the
E-W distance for the T2KA orientation.  A clear systematic error is seen in
both clusters.}
\end{figure}

\begin{figure}
\figurenum{5}
\label{fig:masking}
\caption{The core of Abell~1234, multiplied by the isophotal mask.
From left to right, and top to bottom, the surface brightness limit
$\mu_{limit}$ is 26, 26.5, 27, and 27.5, respectively.  Note that
the brighter surface brightness limits allow some galaxy light
to ``leak'' around the edges of the mask.  See the text for further
discussion.}
\end{figure}

\begin{figure}
\figurenum{6}
\label{fig:fractions}
\plotone{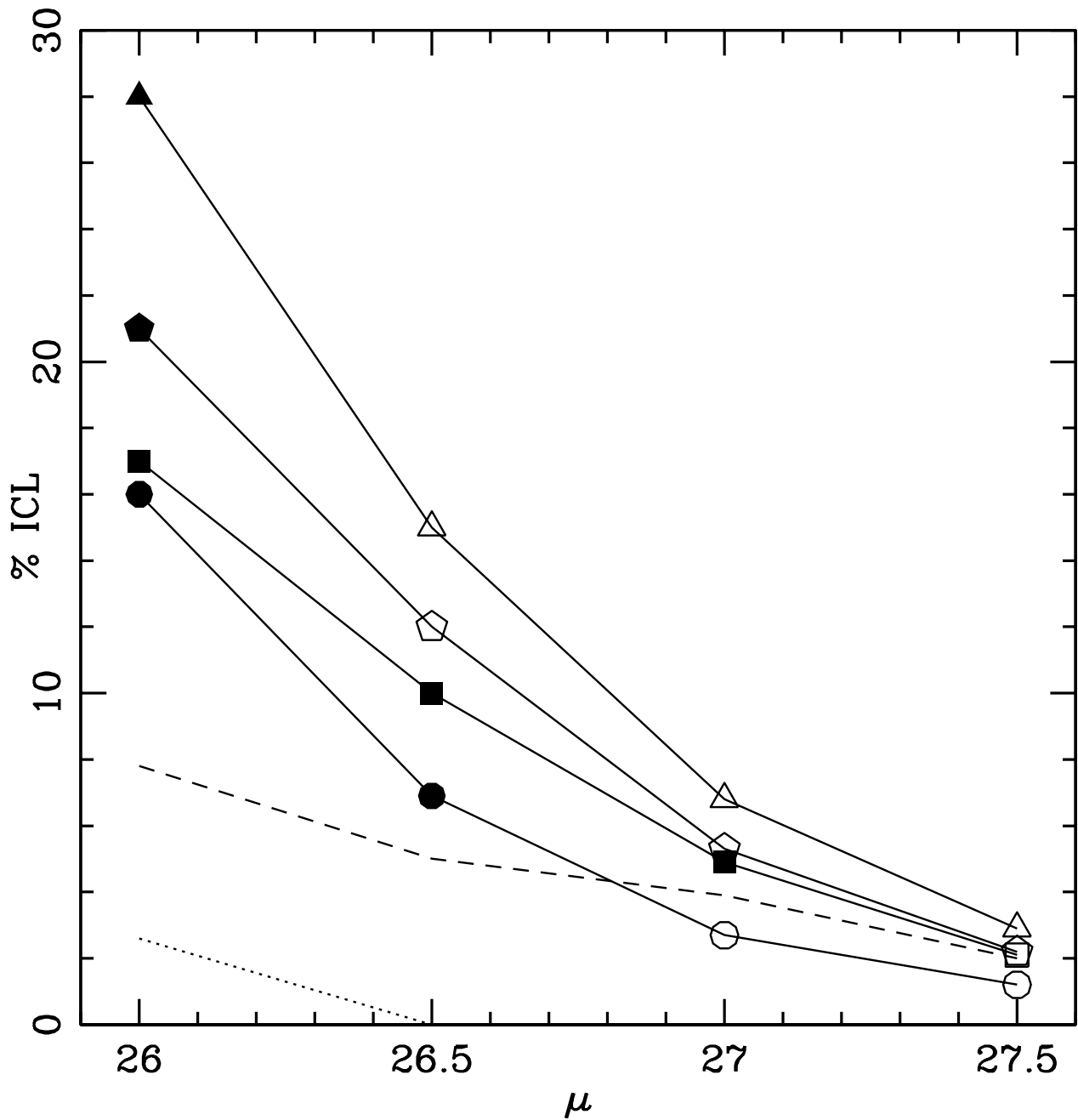}
\caption{A comparison of the derived intracluster luminosity 
fractions for each cluster.  Circles denote data from Abell~801, squares
are from Abell~1234, pentagons are from Abell~1553, and the triangles
denote data from Abell~1914.  Filled points denote measurements 
above $\mu_{5\sigma}$, and open points denote measurements 
below $\mu_{5\sigma}$.  These measurements are compared to models
of the galaxy NGC~3379 (the dashed line), and NGC~3379 tidally
truncated (the dotted line).  See the text for further discussion.
Note that the error bars have been omitted from this 
figure for clarity, but the errors are significant, 
varying from $\approx$ 4\% for Abell 801 and 1234, and up to
$\approx$ 16\% for Abell~1553 and Abell~1914.  The error
bars are given explicitly in Table~5.}
\end{figure}

\begin{figure}
\figurenum{7}
\label{fig:coloricl}
\caption{Images of Abell~801 (top left), and Abell~1234 (top right),
Abell~1553 (bottom left), and Abell~1914 (bottom right binned up into
11 $\times$ 11 pixels bins.  North is left, and east is at the bottom
of these images.  The color black represents all bins with an average
surface brightness from \muv ~= 20 to 24, the color red represents all
bins from \muv ~= 24 to 26, the color green represents all bins from
\muv ~= 26 to 27 and the color blue represents all bins from \muv ~=
27 to 28, which is only reliably reached in the Abell~801 and
Abell~1234.  All bins with surface brightnesses below \muv ~= 28 (for
Abell~801 and Abell~1234) or \muv ~= 27 (for Abell~1553 and
Abell~1914) are left uncolored.  For Abell~801, a flat-fielding
fluctuation can be seen at the top of the image, while for Abell~1553,
and Abell~1914 the systematic large scale flat fielding error
described in Figure~\ref{fig:flaterror} can be seen across the
images. In Abell~1553, the bright compact galaxy pair VIII ZW 192 is
present in the bottom right hand portion of the image (Zwicky,
Sargent, \& Kowal 1975).}
\end{figure}


\begin{figure}
\figurenum{8}
\label{fig:centroid}
\caption{Images of the central regions of each cluster, binned up into
regions of 11 $\times$ 11 pixels (3.3\arcsec).  From left to
right and top to bottom they are Abell~801, Abell~1234, Abell~1553, and
Abell~1914.  North is to the left on each image, and east is at the
bottom.  The centroid is marked with the corresponding surface 
brightness limit.  Abell~801 and 1553 have small deviations in the centroid
position, where Abell~1234, and Abell~1914 have large deviations.  See
the text for further discussion.}
\end{figure}

\begin{figure}
\figurenum{9}
\label{fig:a1553plume}
\caption{Two images of a central portion of Abell~1553.  The images
are both 96\arcsec square (corresponding to $\approx 250$ kpc at our
adopted distance).  North is to the left, and east is at the bottom
of these images.  The greyscale of both images runs from \muv ~= 24.6
(black) to \muv ~= 29 (white).  On the left is the reduced image, 
without any masking or binning.  A plume-like surface brightness 
feature is clearly visible extending from the most luminous galaxy 
towards the southeast.  The surface brightness of this feature is
approximately \muv ~= 25.1.  On the right is the same image, but
binned up into 11 $\times$ 11 pixels, and with the smaller galaxies
masked.  In this figure, the brighter plume is connected with a
much fainter plume of average surface brightness of \muv ~= 26.3.
The small white points denote the approximate extent of this fainter
plume.  See the text for further discussion.}
\end{figure}

\begin{figure}
\figurenum{10}
\label{fig:a1914global}
\caption{A 4.0\arcmin ~by 3.1\arcmin ~image of the central region of 
Abell~1914. North is to the left and east is at the bottom of this
figure.  The five tidal features discussed in the text are labeled.}
\end{figure}

\begin{figure}
\figurenum{11}
\label{fig:westarc}
\caption{A 2.0\arcmin ~by  1.6\arcmin ~image of the western arc of
Abell~1914.  North is to the left, and east is at the bottom of this
figure.  The dark lines outline the arc structure, which is
approximately 1\arcmin ~long, and 12\arcsec ~wide at the widest point.
Note the diffuse structure of the arc, unlike those found from 
gravitational lensing.}
\end{figure}

\begin{figure}
\figurenum{12}
\label{fig:eastplume}
\plotone{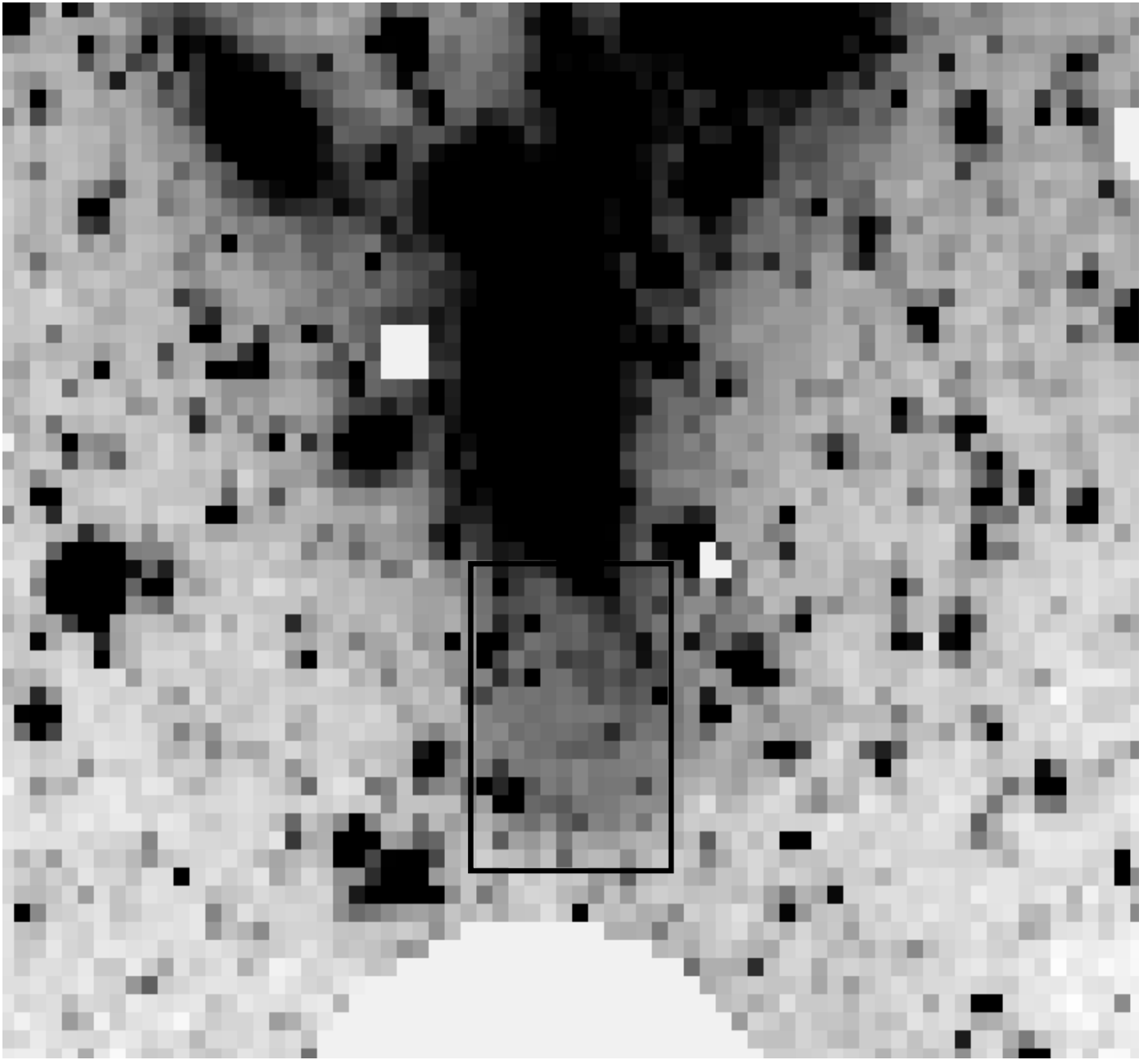}
\caption{A 4.0\arcmin ~by  3.2\arcmin ~image of the eastern plume of
Abell~1914.North is to the left, and east is at the bottom of this
figure.  The dark lines outline the structure of the plume, which
is extremely elongated, and has a steep surface brightness gradient.  
The large white circle at the bottom of this image is the edge of 
a saturated star.}
\end{figure}

\begin{figure}
\figurenum{13}
\label{fig:simimages}
\caption{Images of some of the simulated clusters of Dubinski (1998), colored
in an identical way as in Figure~\ref{fig:coloricl}.
The top row shows a single cluster as it evolves over time, while the
bottom column shows four clusters of different morphologies at z = 0.
Note the asymmetric surface brightness distribution at lower surface
brightnesses, and the range of ICL properties.  See the text for further
discussion.}
\end{figure}

\begin{figure}
\figurenum{14}
\label{fig:simfrac}
\plotone{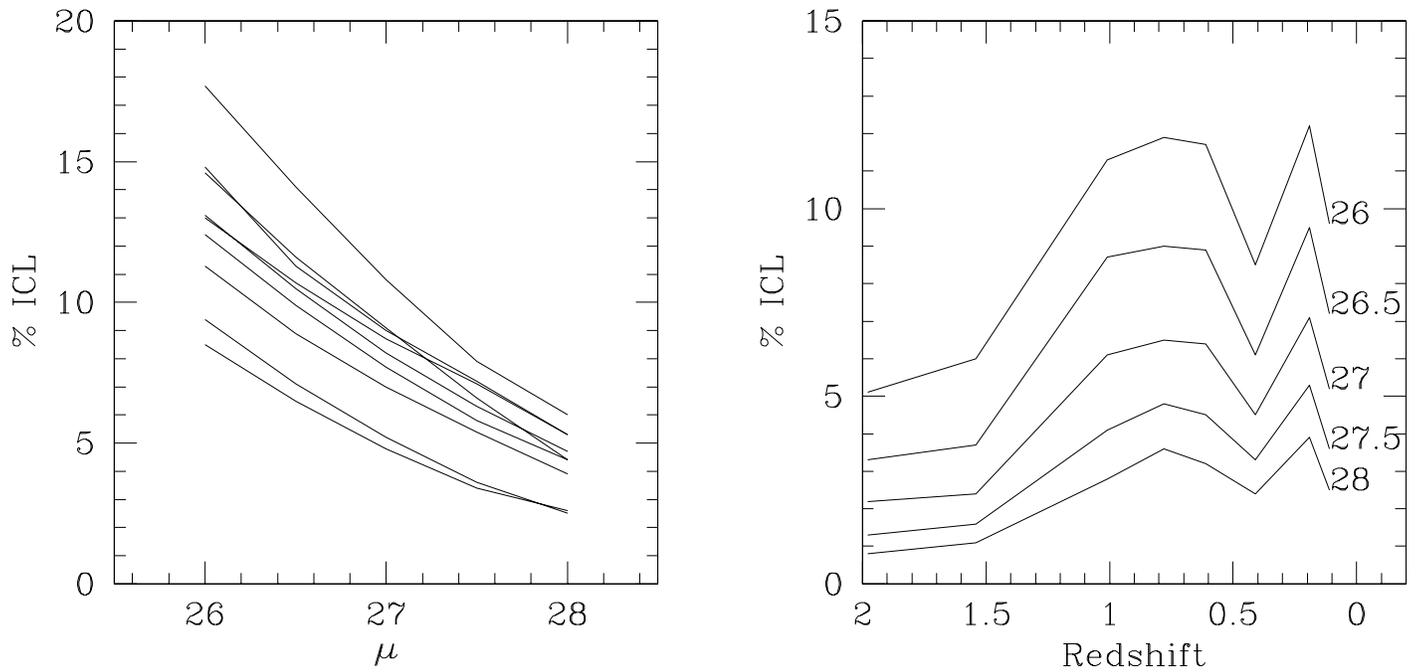}
\caption{The ICL luminosity fraction found in the 
simulations by Dubinski \etal (2003), 
as a function of surface brightness, measured in an identical way as the
observations.  On the left is the results from nine clusters determined
at a redshift of zero.  On the right is the results from an single evolving
cluster as a function of redshift.  See the text for further explanation.}
\end{figure}

\end{document}